***Strain-induced enhancement of the electron energy relaxation in strongly correlated superconductors***

C. Gadermaier[1]*, V. V. Kabanov[1], A. S. Alexandrov[1,2,3], L. Stojchevska[1], T. Mertelj[1], C. Manzoni[4], G. Cerullo[4], N. D. Zhigadlo[5], J. Karpinski[5], Y.Q.Cai[6], X. Yao[6], Y. Toda[7], M. Oda[8], S. Sugai[9,10], and D. Mihailovic[1]

[1]*Department of Complex Matter, Jozef Stefan Institute, Jamova 39, 1000 Ljubljana, Slovenia.*

[2]*Department of Physics, Loughborough University, Loughborough LE11 3TU, United Kingdom.*

[3]*Instituto de Física 'Gleb Wataghin'/DFA, Universidade Estadual de Campinas-UNICAMP 13083-859, Brazil.*

[4]*IFN-CNR, Dipartimento di Fisica, Politecnico di Milano, Piazza L. da Vinci 32, 20133 Milano, Italy.*

[5]*Laboratory for Solid State Physics ETH Zurich, 8093 Zurich, Switzerland.*

[6]*Department of Physics, Shanghai Jiao Tong University, 800 Dongchuan Road, Shanghai 200240, China.*

[7]*Department of Applied Physics, Hokkaido University, Sapporo 060-8628, Japan.*

[8]*Department of Physics, Hokkaido University, Sapporo 060-0810, Japan.*

[9]*Department of Physics, Art and Science, Petroleum Institute, P.O. Box 2533, Abu Dhabi, UAE.*

[10]*Department of Physics, Faculty of Science, Nagoya University, Furo-cho, Chikusa-ku, Nagoya 464-8602, Japan.*

We use femtosecond optical spectroscopy to systematically measure the primary energy relaxation rate $\Gamma_1$ of photoexcited carriers in cuprate and pnictide superconductors. We find that $\Gamma_1$ increases monotonically with increased negative strain in the crystallographic *a*-axis. Generally, the Bardeen-Shockley deformation potential theorem and, specifically, pressure-induced Raman shifts reported in the literature suggest that increased negative strain enhances electron-phonon coupling, which implies that the observed direct correspondence between *a* and $\Gamma_1$ is consistent with the canonical assignment of $\Gamma_1$ to the electron-phonon interaction. The well-known non-monotonic dependence of the superconducting critical temperature $T_c$ on the *a*-axis strain is also reflected in a systematic dependence $T_c$ on $\Gamma_1$, with a distinct maximum at intermediate values (~16 $ps^{-1}$ at room temperature). The empirical non-monotonic systematic variation of $T_c$ with the strength of the electron-phonon interaction provides us with unique insight into the role of electron-phonon interaction in relation to the mechanism of high-$T_c$ superconductivity as a crossover phenomenon.

The deformation potential theorem [1] has been very successful in rationalizing and quantifying the strength of the electron-phonon interaction (EPI) in relation to charge carrier scattering and mobility in semiconductors [2,3], thus rapidly leading to an understanding of charge carrier dynamics and opening the way to modern semiconductor technology. Subsequently, it has been shown that the formalism is not limited to semiconductors, but can be extrapolated both to metals and insulators [4]. Strain on the structure caused by external pressure or by doping may result in significant changes of the EPI, leading to changes in functional properties over and above those caused by the changes of density of doped charges.

In both the cuprate and pnictide families of superconductors, the inter-atomic distance has been discussed as a quantitative parameter [5-7] besides the doping level $x$ that is systematically correlated with the superconducting critical temperature $T_c$. Although this was pointed out in cuprates long ago and is obviously of primary relevance to the superconducting mechanism, there is not even an elementary understanding on the origin of the peculiar non-monotonic variation of $T_c$ on the lattice constant which is ubiquitously observed. Here we present systematic measurements of electron energy relaxation in the normal state of different high $T_c$ superconductors to show that in both cuprate and pnictide superconductors the primary electron energy relaxation rate $\Gamma_1$ is directly correlated with the length $a$ of the crystallographic $a$-axis, which in turn depends on the structure, the type of dopant atoms and the doping level $x$. The dependence of the deformation potential on the lattice constant rationalizes how the EPI strength increases with increasing negative strain in the $a$-axis. Using established models which relate $\Gamma_1$ to the EPI, we make the connection between the EPI and $T_c$ for the first time. From the observed characteristic variation of $T_c$ on the EPI we can reach important conclusions regarding the relevance of EPI to the superconducting mechanism in the context of strong correlations.

Following the proposal [8] and demonstration [9] of the use of femtosecond optical pump-probe spectroscopy to investigate the EPI in classical superconductors, we use an ultrashort laser pulse (the pump) to create a non-equilibrium electron distribution. The electron energy relaxation is probed by a second, weaker probe pulse, which detects the transient change in the reflectivity $\Delta R/R$. This method has been shown to yield a more reliable measure of the EPI strength[9-14] than the measurement of phonon linewidths by Raman or neutron scattering, which are often compromised by inhomogeneous broadening and probe only a limited selection of phonons.

Although the technique is experimentally well established, some precautions need to be taken to obtain quantitatively accurate data on $\Gamma_1$. In order to disentangle the various relaxation pathways in high-$T_c$ superconductors and determine their time constants $\tau_i = 1/\Gamma_i$ we use several different probe wavelengths. Since the $\tau_i$ depend on the sample temperature (See Equation 1), and additional processes appear in the presence of low-temperature order [15], we measure all samples at the same temperature - 300K - well above the temperature where any low-temperature ordering might take place. We avoid non-linear, intensity dependent relaxation processes and sample heating or degradation by keeping the pump pulse fluence below 20 $\mu Jcm^{-2}$. Most importantly, to provide the necessary time resolution and sensitivity, we purpose-built a system that gives tunable pump and probe wavelengths ranging from 500 to 700 nm with sub-30-fs overall time resolution. Sub-25-fs pump pulses with a repetition rate of 250 kHz are provided by a non-collinear optical parametric amplifier tuned to a center wavelength of 535 nm (see the Supplementary Information for a detailed description). Probe pulses are provided by a white light continuum generated in 2.5 mm thick sapphire. Adequate time resolution is ensured via spectral filtering of the chirped probe pulses [16,17]. In several of the samples investigated here, this allowed the identification of previously unresolved relaxation processes.

The sample growth and doping is described elsewhere [18-23]. Samples are glued to a copper support using GE Varnish. When needed, samples are cleaved to obtain a reflecting surface with low scattering. The reflectivity and the relative spectral weight of the signal components may vary across the surface, but the relaxation times do not. To eliminate the influence of the sample thickness we only used samples much thicker than 100 μm, while our experiment probes a surface layer typically ~100 nm, depending on the material. The superconducting critical temperature $T_c$ (see Table 1) was determined by measuring the temperature dependence of the AC magnetic susceptibility using a superconducting quantum interference device magnetometer.

A representative data set on the transient reflectivity of the selected compounds at room temperature is shown in Fig. 1. (The complete data are reported in the Supplementary Information.) To extract the relaxation times, we fit the data using single- or double-exponential decays and a Gaussian function for the instrumental response. (See the Supplementary Information on the time resolution of the used setup and its implications for the accuracy of short $\tau_1$ values). Oscillatory components are modelled using the same generation term as the main signal, and leaving as fit parameters amplitude, frequency, phase, and damping time. We measure all samples over a range of different wavelengths and in both orthogonal polarisations. Although we let all fit parameters vary freely between different wavelengths, for each sample the best fits yield the same relaxation times within 10 - 20 %. From this observation we conclude that at different wavelengths we are measuring the same relaxation processes, but with different spectral weights.

From the fits we find that the primary (fastest) relaxation rate $\Gamma_1$ ranges from 3 to 25 ps$^{-1}$ (i. e. relaxation time $\tau_1$ between 40 and 330 fs, Table 1), independent of probe wavelength and sample orientation with respect to the probe polarisation. For the cuprates we also identify a slower relaxation component $\tau_2$. Some of the transient reflectivity signals show strong coherent phonon oscillations [24], which are subtracted in the analysis. The $\tau_1$ values for $YBa_2Cu_3O_{6+x}$, $Bi_2Sr_2CaCu_2O_{8.14}$, and the pnictides agree well with room temperature data from the literature [10-13], while for $HgBaCa_2CuO_{4.1}$ they are resolved for the first time. For $La_{2-x}Sr_xCuO_4$ our recent results obtained with much higher pump intensity [14] are confirmed by the present low intensity data.

In Fig. 2 we plot the primary relaxation rate $\Gamma_1$ as a function of the crystallographic *a*-axis of pnictides and cuprates. All data points for cuprates, including differently doped samples, fit to a straight line, where $\Gamma_1$ decreases with increasing *a*. Defining the *a*-axis strain $\varepsilon_a = \Delta a/a_0$, with the equilibrium value [6] $a_0 = 3.94$ Å, this means that higher negative strain in the Cu-O planes leads to faster electron energy relaxation. For a given cuprate, *a* is not determined only by the type of compound, but also varies significantly with doping, by 0.1 Å over the whole superconducting range of the phase diagram [5]. This translates into a doping dependent relaxation rate, as our data for $YBa_2Cu_3O_{6+x}$ and $La_{2-x}Sr_xCuO_4$ at different doping levels confirm, and is in agreement with previous data [11] on $Bi_2Sr_2CaCu_2O_{8+x}$. On the other hand [25], *a* in the pnictide $Ba(Fe_{1-x}Co_x)_2Se_2$ is nearly doping independent, changing by less than 0.002 Å. Remarkably, for all measured $Ba(Fe_{1-x}Co_x)_2Se_2$ samples, which cover a wide range of doping levels, the measured $\Gamma_1$ values show almost no variation. Hence $\Gamma_1$ depends on *a*, but not on the doped carrier density.

Given the consensus in the literature in attributing the ultrafast response $\Gamma_1$ to the electron energy relaxation [10-14,26-28], we proceed to discuss the relaxation mechanism. The electron-phonon relaxation rate $1/\tau_{e-ph}$ can be related to the EPI strength expressed as the second moment $\lambda\langle\omega^2\rangle$ of the Eliashberg spectral function [8,29]:

$$\frac{1}{\tau_{e-ph}} = B\frac{\lambda\langle\omega^2\rangle}{T} \qquad (1)$$

where $T$ is the temperature and $B$ is a constant. Depending on whether one assumes the electron-electron relaxation to be much faster than the electron-phonon relaxation [8], or not [29], $T$ is either the electronic temperature $T_e$ or the lattice temperature $T_L$, respectively. In the present experiments we use a very low laser excitation fluence, in which case $T_e \sim T_L$. The models above assume coupling between two baths: the hot electrons and the phonons whose coupling is characterized by the Eliashberg spectral function. One may also include more specific bosonic excitations in such models to emphasize a particular interaction [11,26-28]. However, this does not have significant bearing in the current discussion, since we are discussing only the dominant relaxation rate. The electron energy relaxes towards the bath which has the largest heat capacity; the lattice is the obvious candidate [30]. Recent microscopic calculations within the driven *t*-*J*-Holstein model suggest that under certain conditions rapid energy transfer to spin degrees of freedom can occur [31], but considering the relaxation dynamics, only the relaxation of a highly-excited charge carrier through Holstein phonons has been investigated so far [32]. A detailed discussion of the assignment of $\Gamma_1$ to the EPI and the choice of the proportionality constant $B$ is given in the supplement.

The EPI strength is often described by a dimensionless parameter $\lambda$ (the zero-order moment of the Eliashberg spectral function) rather than $\lambda\langle\omega^2\rangle$. To calculate $\lambda$ from $\lambda\langle\omega^2\rangle$, one would need to know the complete $q$- and $\omega$-dependent Eliashberg spectral function. As a crude approximation, $\lambda\langle\omega^2\rangle$ is often simply divided by the square of an effective phonon frequency $\omega^*$. Commonly used values are 40 meV for cuprates [20,28] and 25 meV for pnictides [23]. Using these values, we estimate $\lambda \sim 0.3 \div 0.5$ for cuprates and $\lambda \sim 0.2 \div 0.3$ for pnictides. This estimate agrees well with $\lambda$ obtained from angle-resolved photoemission [33,34], neutron scattering [35-37], and tunnelling [38] spectroscopy.

Generally, the effective electron-phonon coupling $\lambda$ is proportional to the square of the energy $E_1$, defined via the deformation potential [1,39] $\delta U = E_1 \Delta$, which gives the energy shift of the conduction band edge as a function of the strain $\Delta$. For a small variation of $a$ (for the cuprates $\Delta a/a \sim 3\%$), we may expand around the equilibrium value $a_0$ to first order:

$$\lambda^*(a) \approx \gamma \left[ E_1(a_0) + \frac{\partial E_1}{\partial a}(a - a_0) \right]^2 \approx \lambda^*(a_0) + \tilde{\gamma} \frac{\partial E_1}{\partial a}(a - a_0) \qquad (2)$$

The linear relation between $\lambda$ and $a$ above justifies the fit of $\Gamma_1(a)$ in Fig 2. Using a simple estimate for $E_1 \approx \hbar^2 / 3ma^2$, where $m$ is the electron mass, $\partial E_1 / \partial a$ is negative, in agreement with the negative slope of the experimentally obtained $\Gamma_1(a)$.

In addition to the increase of $\lambda$ with decreasing $a$, also the phonon frequencies observed by Raman spectroscopy increase [40,41]. Both effects lead to an increase of the relaxation rate (see next paragraph). The observed blue-shifts of the phonon bands by a few % per pm of contraction of $a$ are consistent both in sign and magnitude with the observed increase of $\Gamma_1$.

In both cuprates and pnictides it has been proposed that $T_c$ is determined by two parameters: the doping level $x$ and the length $a$ of the crystallographic $a$-axis [5-7], or equivalently some related structural parameter, such as the Cu-O in-plane bond length or the anion height in the pnictides. In the cuprates, superconductivity occurs when structure, doping- or pressure-induced strain compresses $a$ well below its equilibrium value [5,6] (in a cubic lattice) of $a$ = 3.94 Å, with a maximum $T_c$ at $a$ = 3.84 Å (Fig. 3a). The direct correspondence between $a$ and $\Gamma_1$ shown in Figure 2 suggests that, equivalently to $a$, $\Gamma_1$ can be used as a second parameter which determines $T_c$ (see Fig. 3b). $T_c(R)$ systematically follows an arc with a distinct maximum around $k \sim 16$ ps$^{-1}$. For the cuprates, $a$ and consequently $\Gamma_1$ depend on the doping levels, while for $Ba(Fe_{1-x}Co_x)_2Se_2$ the two parameters are practically independent. The femtosecond pump-probe experiment thus directly yields a parameter - $\Gamma_1$ - that is uniquely related to $T_c$.

Allen's proposition [8] that $\Gamma_1$ be proportional to the EPI and the Bardeen-Shockley deformation potential theorem [1] that relates the EPI to strain on the *a*-axis together suggest a correspondence between *a* and the EPI, which we verified experimentally. Relating $T_c$ to the EPI rather than to *a* can give us better insight into the applicability of different mechanisms of superconductivity. BCS theory, and by extension Eliashberg theory, predict a monotonically increasing $T_c$ with increasing $\lambda$, which is contrary to the systematics we observe. On the other hand, in bare strongly correlated electron models, such as the Hubbard model and the *t*-*J* model, the tunneling *t* decreases with increasing distance, but the dependence of $T_c$ on *t* is not clear. Until now, the dependence of other interactions, such as spin fluctuations, on the *a* parameter has not been studied, but it is possible that these may have an – as yet undetermined – effect on $T_c$.

The observed non-monotonic behavior strongly suggests a crossover phenomenon. Thus competing interactions need to be considered, such as the crossover from weak coupling (BCS-like) to strong coupling [42] as a function of $\lambda$. An alternative approach includes strong correlations via the Coulomb interaction into an EPI model within the strong coupling approach [43]. In other words, the crossover behavior also highlights the necessity of the interplay of competing ground states [19] for achieving high $T_c$s. In any case the EPI is clearly involved in a very unconventional way [44]. Indeed, the systematic, large, yet strongly unconventional isotope effect departing strongly from BCS behavior observed across the compounds corroborates this notion [45,46]. Although the present data and analysis presented in Fig. 3 do not identify the mechanism, they emphasize a role for unconventional electron-lattice effects and give a stringent verification criterion for any hopeful high-$T_c$ superconductivity theory [47,48].

We thank D. Brida, F. Cilento, I. R. Fisher, P. Kusar, D. Polli, and L. Vidmar for stimulating discussions. This work was supported by the Slovenian Research Agency (ARRS) (grants 430-66/2007-17, BI-CN/07-09-003, and BI-IT/11-13-001), ROBOCON 2011-2012, the Royal Society (grant JP090316), and by the European Commission (grant ERG-230975 and the European Community Access to Research Infrastructure Action, Contract RII3-CT-2003-506350 (Centre for Ultrafast Science and Biomedical Optics, LASERLAB-EUROPE)).

Figure Captions:

Figure 1. Transient differential reflection at 295 K for different (near) optimally doped superconducting compounds. Thin solid lines are fits. Please refer to the supplement for different doping levels and sample orientations (Figures S8-S19).

Figure 2. The primary electron energy relaxation rate $\Gamma$ of pnictides and cuprates as a function of the $a$-axis length. Large symbols are at or close to optimal doping, open symbols at other doping levels. Dashed lines are linear fits. The inset is a zoom into the region of the five different $Ba(Fe_{0.975}Co_{0.025})_2As_2$ samples, which have very similar a and $\Gamma$ values.

Figure 3. a) Critical temperature for cuprates as a function of the $a$-axis (two times the in-plane Cu-O distance as published in Ref. 5 - please refer to the caption of Fig. 6 of that manuscript for the numbering of the compounds). Full symbols are our samples at or close to optimal doping (same symbol/color coding as Fig. 2), open circles are from Ref. 5, line is a fit to a Lorentzian as a guide to the eye. b) The superconducting transition temperature $T_c$ as a function of the primary relaxation rate $\Gamma_1$. For plotting the top axis we have used Equation 1 with[33] $B = 3\hbar/2\pi k_B$. Full symbols indicate (almost) optimally doped samples, open symbols other doping levels.

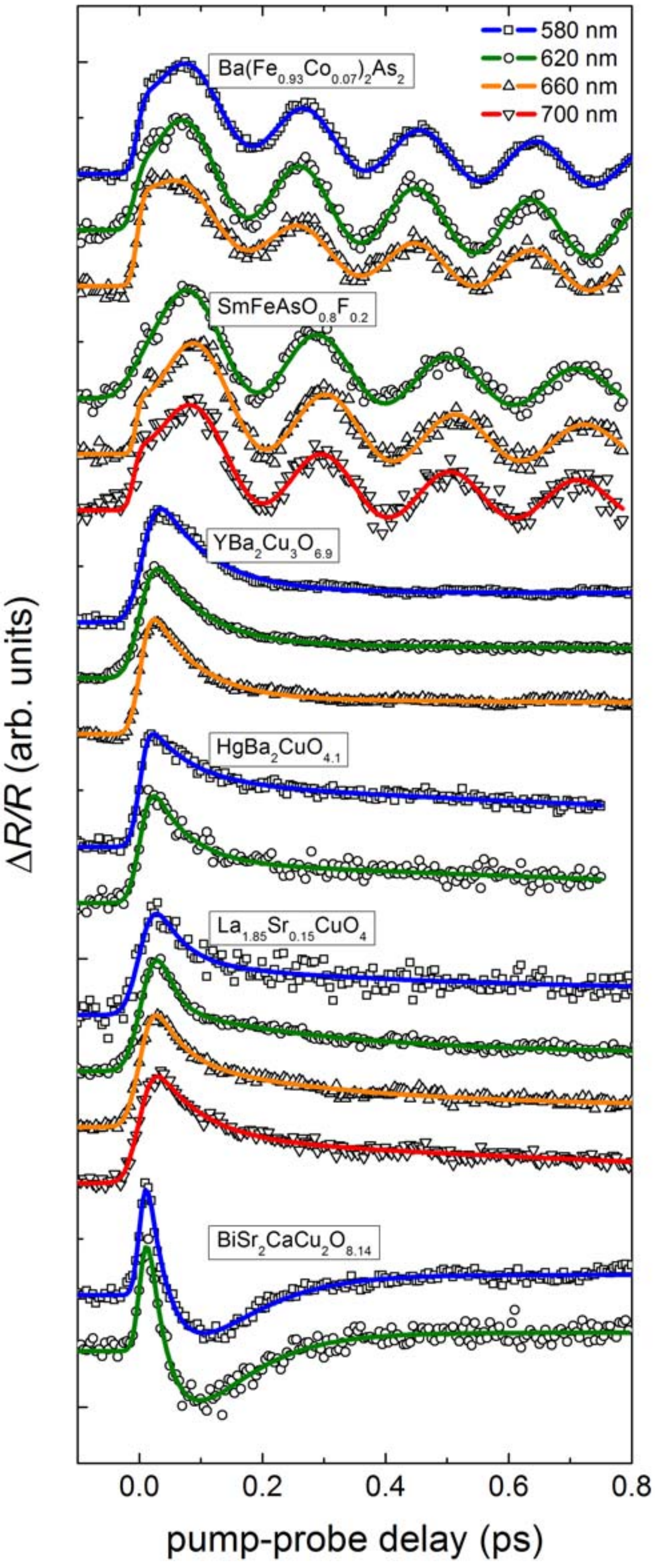

C. Gadermaier et al. Fig. 1

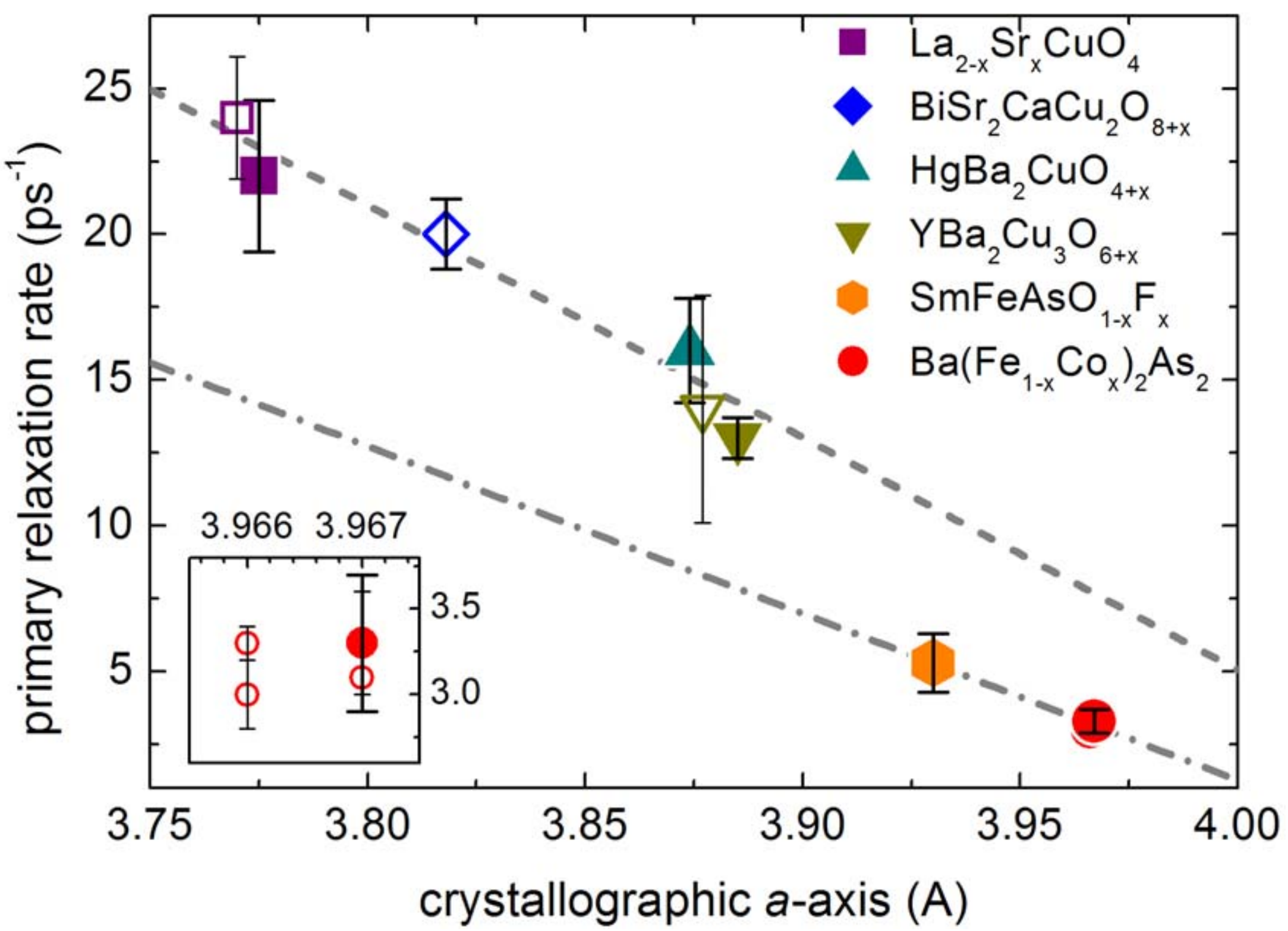

C. Gadermaier et al. Fig 2

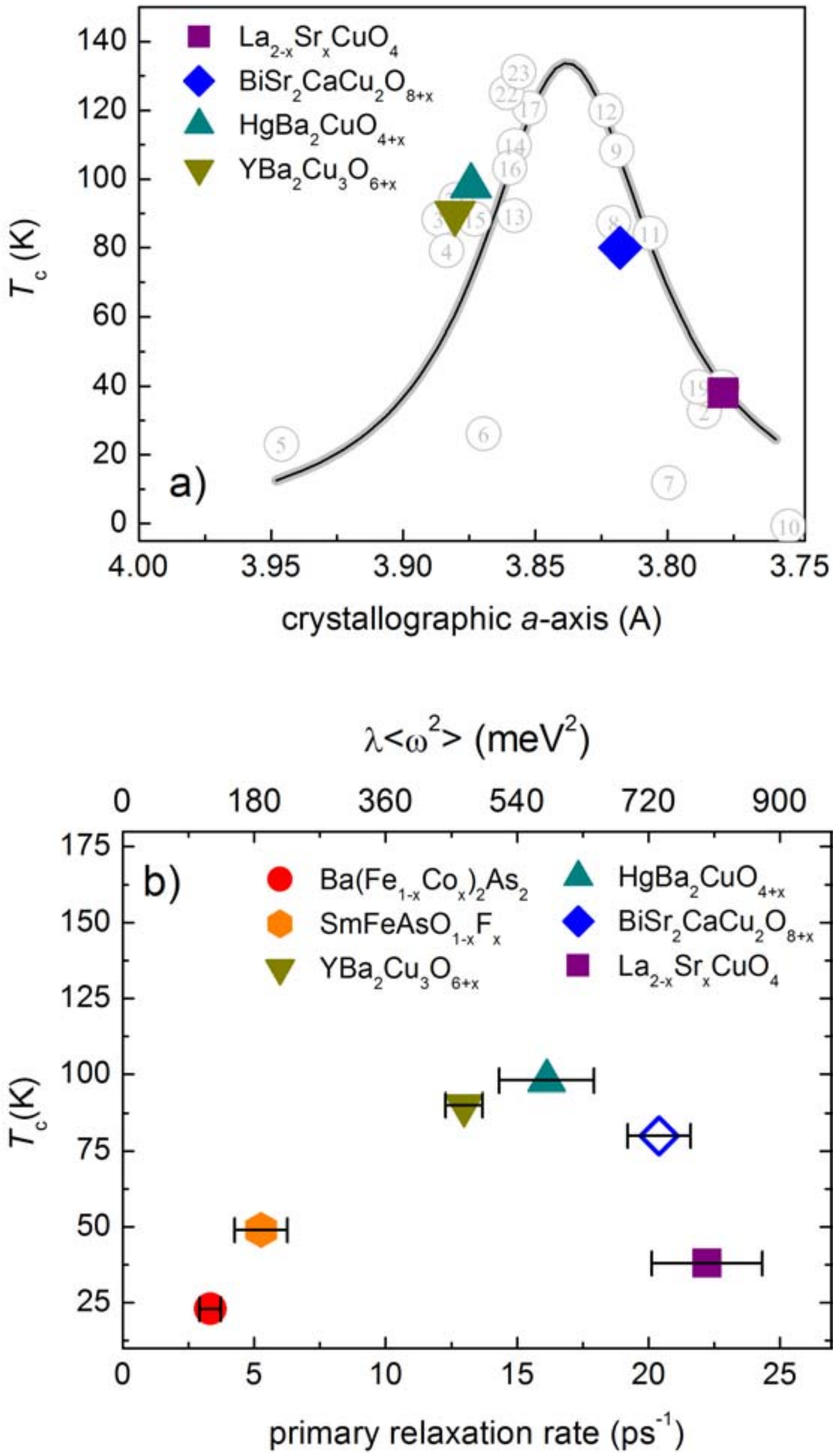

C. Gadermaier et al. Fig 3

Table 1. Stoichiometries, superconducting critical temperatures, and electron-phonon relaxation times and rates of the investigated samples. For each compound family we calculated error margins (standard deviation over the obtained fit parameters) for the sample with the highest number of measured time treasures.

| Sample | $T_c$ (K) | $\tau_{e\text{-}ph}$ (fs) | $1/\tau_{e\text{-}ph}$ ($ps^{-1}$) |
|---|---|---|---|
| $BaFe_2As_2$ | 0 | 300 | 3.3 |
| $Ba(Fe_{0.975}Co_{0.025})_2As_2$ | 0 | 330 | 3.0 |
| $Ba(Fe_{0.949}Co_{0.051})_2As_2$ | 20 | 320 | 3.1 |
| $Ba(Fe_{0.93}Co_{0.07})_2As_2$ | 23 | 300 | 3.3 |
| $Ba(Fe_{0.89}Co_{0.11})_2As_2$ | 10 | 300 | 3.3 |
| $SmFeAsO_{0.8}F_{0.2}$ | 49 | 190 | 5.3 |
| $YBa_2Cu_3O_{6.5}$ | 63 | 72 | 14 |
| $YBa_2Cu_3O_{6.9}$ | 90 | 77 | 13 |
| $HgBa_2CuO_{4.1}$ | 98 | 62 | 16 |
| $Bi_2Sr_2CaCu_2O_{8.14}$ | 80 | 49 | 20 |
| $La_{1.9}Sr_{0.1}CuO_4$ | 30 | 42 | 24 |
| $La_{1.85}Sr_{0.15}CuO_4$ | 38 | 45 | 22 |

*Please contact C. G. at christoph.gadermaier@ijs.si for correspondence and reprint requests.

# Supplementary Material: Strain-induced enhancement of the electron energy relaxation in strongly correlated superconductors

C. Gadermaier[1]*, V. V. Kabanov[1], A. S. Alexandrov[1,2,3], L. Stojchevska[1], T. Mertelj[1], C. Manzoni[4], G. Cerullo[4], N. D. Zhigadlo[5], J. Karpinski[5], Y.Q.Cai[6], X. Yao[6], Y. Toda[7], M. Oda[8], S. Sugai[9,10], and D. Mihailovic[1]

[1]*Department of Complex Matter, Jozef Stefan Institute, Jamova 39, 1000 Ljubljana, Slovenia*

[2]*Department of Physics, Loughborough University, Loughborough LE11 3TU, United Kingdom*

[3]*Instituto de Física 'Gleb Wataghin'/DFA, Universidade Estadual de Campinas-UNICAMP 13083-859, Brazil*

[4]*IFN-CNR, Dipartimento di Fisica, Politecnico di Milano, Piazza L. da Vinci 32, 20133 Milano, Italy*

[5]*Laboratory for Solid State Physics ETH Zurich, 8093 Zurich, Switzerland*

[6]*Department of Physics, Shanghai Jiao Tong University, 800 Dongchuan Road, Shanghai 200240, China*

[7]*Department of Applied Physics, Hokkaido University, Sapporo 060-8628, Japan.*

[8]*Department of Physics, Hokkaido University, Sapporo 060-0810, Japan.*

[9]*Department of Physics, Art and Science, Petroleum Institute, P.O. Box 2533, Abu Dhabi, UAE*

[10]*Department of Physics, Faculty of Science, Nagoya University, Furo-cho, Chikusa-ku, Nagoya 464-8602, Japan*

## A. Methods

A1 femtosecond pump-probe spectroscopy:

In femtosecond optical pump-probe spectroscopy a first laser pulse (the pump) creates a non-equilibrium electron energy distribution and a second (weaker) pulse probes the change in the reflectivity $R$ or transmittivity $T$ of the sample as a function of pump-probe delay $t_{PP}$. The relative change $\Delta R/R(t_{PP})$ or $\Delta T/T(t_{PP})$ directly tracks the electronic relaxation processes[48,48]. In order to yield useful data, a pump-probe experiment must meet the following requirements: (i) provide sufficient time resolution to track all relevant processes, (ii) allow reliable disentanglement of the different processes, (iii) avoid sample degradation and heating, and (iv) provide a good compromise between the signal-to-noise ratio and the time needed for alignment and data acquisition. The time resolution is given by the cross-correlation between the pump and probe pulse intensity profiles if both are transform-limited, i.e. as short as their spectral width $\Delta\omega$ allows ($\tau \sim 1/\Delta\omega$). Non-transform limited pulses are chirped, i.e. different wavelengths are delayed with respect to each other. The time-resolution of a pump-probe experiment using chirped pulses can be improved compared to the cross-correlation by spectral filtering[48,48] of the probe, which can be realized by placing a wavelength selective filter after the sample; selecting a spectral slice of the probe pulse for detection also selects a temporal slice of the chirped pulse. This way the time resolution can be almost as high as for transform-limited pulses of the same spectral width as the unfiltered probe pulse[**Error! Bookmark not defined.**,4]. We use nearly transform-limited pump pulses of about 20 fs duration, modulated at ~1.5 kHz with a mechanical chopper, and broadband probe pulses with partially compensated chirp and with a monochromator placed between the sample and detector (photodiode using phase-sensitive detection with a lock-in amplifier), giving sub-30 fs time resolution. The occurrence of relaxation times as short as 40 fs shows that the high time resolution was crucial in enabling our study.

Since the relaxation of excited electrons proceeds via different processes, the time traces $\Delta R/R(t_{PP})$ are never simple single-exponential decay curves, even if the individual relaxation processes are. Their contribution to the signal varies with wavelength, therefore a probe pulse covering a broad spectral range (i.e. having a broad spectrum and/or tunable centre wavelength) is needed to capture and disentangle all processes involved.

In order to avoid sample degradation, heating, or intensity-dependent non-linear relaxation phenomena, we kept the excitation density below 20 $\mu Jcm^{-2}$. This requires averaging over a large number of pulses, hence we used a 250 kHz set-up and a custom built non-collinear optical parametric amplifier (NOPA), based on the design in Ref 48, which is able to run at such low pump fluences (see section A2 for a detailed description).

A2 the high-repetition rate non-collinear optical parametric amplifier

We use 2.5 μJ, 50 fs pulses centered at 800 nm at a repetition rate of 250 kHz generated by a regenerative amplifier (Coherent RegA 9000). About 10% of the pulse energy is used to generate the white light continuum of the probe beam and another 10% to generate the white light continuum to seed the NOPA. The remaining 2 μJ are frequency doubled to pump the NOPA with approximately 750 nJ pulse energy at 400 nm. Unlike for the typical 1 kHz NOPAs that run on much higher pump energies, here the β-barium borate (BBO) crystal is placed in the Rayleigh range of the focusing lens to achieve the necessary pump intensities. Pump and seed beams overlap non-collinearly in the 2 mm BBO crystal cut for type I phase matching with their propagation directions inside the crystal forming an angle of 3.7°. At this angle phase matching is ensured over a large bandwidth and the bandwidth of the output is given by the temporal overlap between the pump and the chirped seed pulse. Our NOPA yields pulses of approximately 60 nm bandwidth, tunable between centre wavelengths of 500 to 650 nm by changing the pump-seed delay. At 535 nm centre wavelength, which we use for our experiments, the amplified pulse energy is 20 to 30 nJ. A prism compressor compensates most of the chirp and leads to pulse durations around 20 fs.

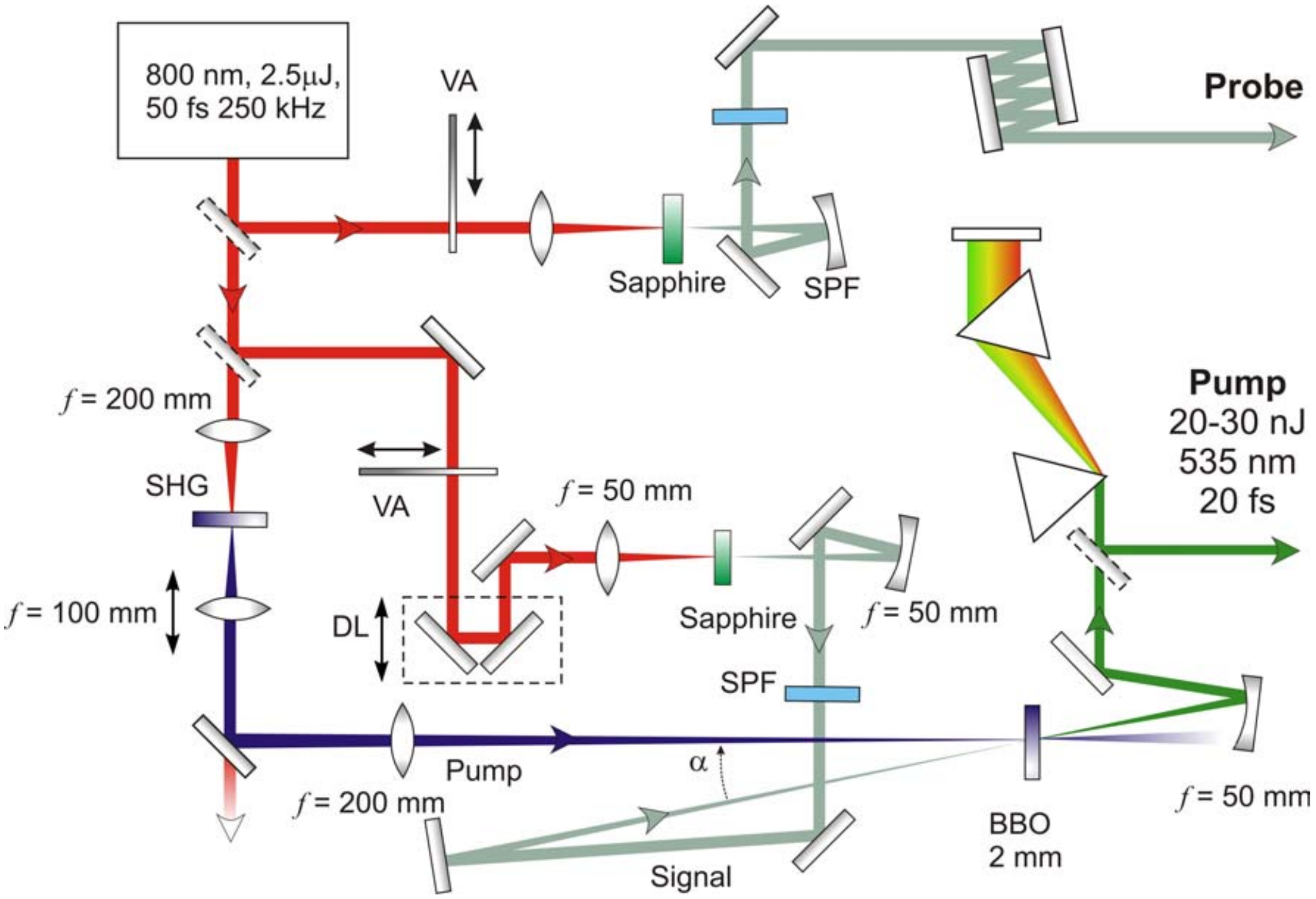

Figure S1. The high-repetition rate NOPA set-up.

A3 The fitting procedure

To extract the relaxation times, we fit the data to the following model curve: (i) As the generation term we use a Gaussian, whose duration is a fit parameter. The duration is assumed to be that of the convolution of the pump intensity profile with the transform limit of the probe intensity profile or slightly longer (see section A2), which we find confirmed in all our data. (ii) Relaxation processes with time constants 30 fs – 3 ps are modeled as single- or double-exponential decays. (iii) slower relaxation processes are modeled as a plateau. (iv) oscillatory components are modeled with the same generation term as the main signal, and leaving as fit parameters amplitude, phase, frequency, and damping time. We measure all samples at different wavelengths (however, due to wavelength dependent reflectivity and signal strengths we do not always have a useful signal at all wavelengths we try) and two perpendicular orientations. Although we let all fit parameters vary freely between different wavelengths, for the same sample the best fits for the relaxation times typically vary only within 10 - 20 %. From this observation we conclude that at different wavelengths we look at the same relaxation processes, only with varying spectral weight. Below we show three examples of data and fits. Note that if the two contributions with $\tau_1$ and $\tau_2$ have opposite sign, as is the case for $Bi_2Sr_2CaCu_2O_{8+x}$, then the fast decay appears faster when the two decay times are similar. Hence the fast decay in the $Bi_2Sr_2CaCu_2O_{8+x}$ signal looks faster than one would expect from the $\tau_1$ value and faster than the signal of $La_{2-x}Sr_xCuO_4$, which has a shorter $\tau_1$.

A4. Assessment of the main sources of uncertainty in the fit

The above described fitting procedure is implemented as coupled differential equations. Although the routine is based on $\chi^2$-sum minimization, it does not yield error margins for each parameter individually. Therefore, in similar studies in the literature it is not customary to report error margins for the observed time constants. Here, we discuss the main sources of uncertainty and how they influence the fit values for $\tau_1$: (i) the instrument resolution, (ii) the consideration of an oscillating component, (iii) the mutual influence of $\tau_1$ and $\tau_2$ on each other, and (iv) the noise of the data.

As discussed above, spectral filtering of the probe light after the sample allows nearly transform-limited instrument resolution. The exact resolution can vary slightly with wavelength and the sample response. Due to the low pump power, the NOPA is running at its lower limit of stability and needs a lot of day-to-day optimization. This causes variations in the pump pulse duration, which was measured by autocorrelation every day and after every major realignment. By fitting the autocorrelation as a Gaussian we obtained pump pulse durations (FWHM) ranging from 20 to 27 fs. Fitting with a hyperbolic secant typically gives 2-3 fs shorter pulses and slightly lower $\chi^2$-sums. The bandwidth of the probe is much larger than that of the pump; hence the best obtainable instrument response $H$ is slightly longer than the pump duration, which may thus serve as a lower limit $H_{min}$. We now assess how this uncertainty affects the fastest measured time constant, $\tau_1$ of $La_{1.9}Sr_{0.1}CuO_4$. The pump duration for the measurements on this sample was $H_{min}$ = 25 fs (from the Gaussian fit to the autocorrelation); the best fit for the dynamics was obtained assuming a generation term with $H_{best}$ = 30 fs duration. We may take $\Delta H = H_{best} - H_{min}$ as an upper estimate for the uncertainty of the instrument response. The fit yields $\tau_1$ = 46 fs for an assumed instrument response of $H_{min}$ and $\tau_1$ = 42 fs for $H_{best}$. Hence the uncertainty in the instrument response at most introduces an uncertainty of $\Delta\tau_1$ = 4 fs, or ~10%. For all other compounds, which have longer $\tau_1$, the uncertainty due to the instrument resolution is less significant.

---

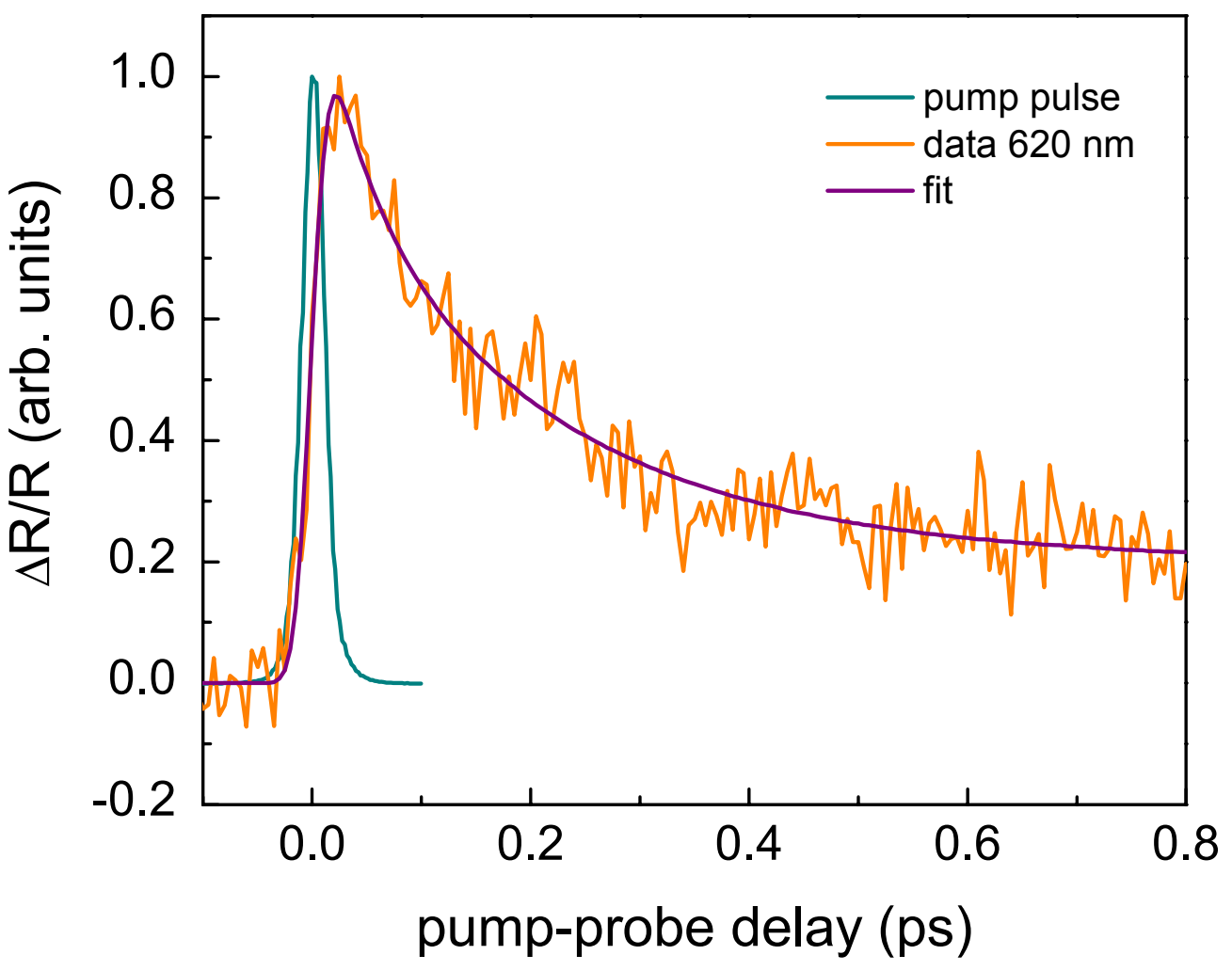

Figure S2. Normalized transient differential reflection of $La_{1.9}Sr_{0.1}CuO_4$ at 620 nm and best fit, assuming a Gaussian generation term with 30 fs duration (the other fit with 25 fs would be indistinguishable in this plot and is not shown). Also shown is an approximate reconstruction of the pump pulse (autocorrelation with the *t*-axis rescaled by a factor $1/\sqrt{2}$, as would be exact for a Gaussian).

Next, we assess how the consideration of the dominant oscillation affects the measured $\tau_1$ for Ba122 and YBCO. Fig S3 shows the data for YBCO6.5 fitted with and without an oscillating mode. With the oscillating contribution we obtain $\tau_1$ = 72 fs and an instrument response of 34 fs. Neglecting the oscillating component yields $\tau_1$ = 77 fs, but an unrealistic instrument response of 64 fs. The peculiar shape of the $\Delta R/R$ signal during the first 200 fs hence gives the illusion of a much slower instrument response if the oscillating component is neglected.

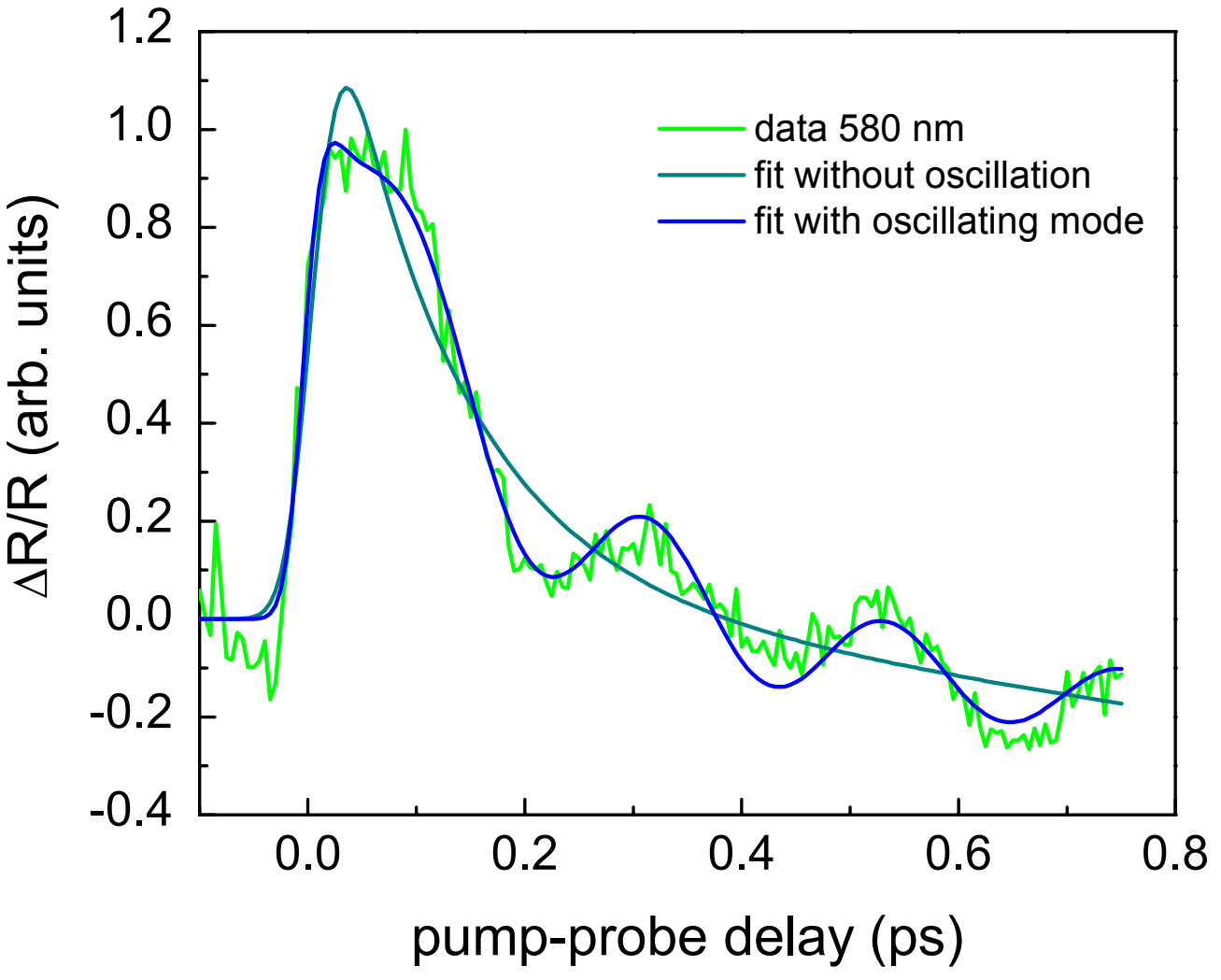

Figure S3. Normalized transient differential reflection of $YBa_2Cu_3O_{6.5}$ at 580 nm and best fits without oscillations and with one oscillating mode.

To evaluate the cross-talk between the fitted $\tau_1$ and $\tau_2$ values for the extreme case of $Bi_2Sr_2CaCu_2O_{8.14}$, where the two signal components have opposite sign and rather similar relaxation times ($\tau_2/\tau_1 < 2$), we fit the time trace at 620 nm with four different values of $\tau_2$ = 50, 75, 100, and 125 fs. We obtain best fits for $\tau_1$ = 50, 49, 38, and 32 fs. The two central values give a visibly better fit (See Figure S4) and can hence be used to estimate $\Delta\tau_1$ = 5 fs, or $\Delta\tau_1/\tau_1 \sim 10\%$.

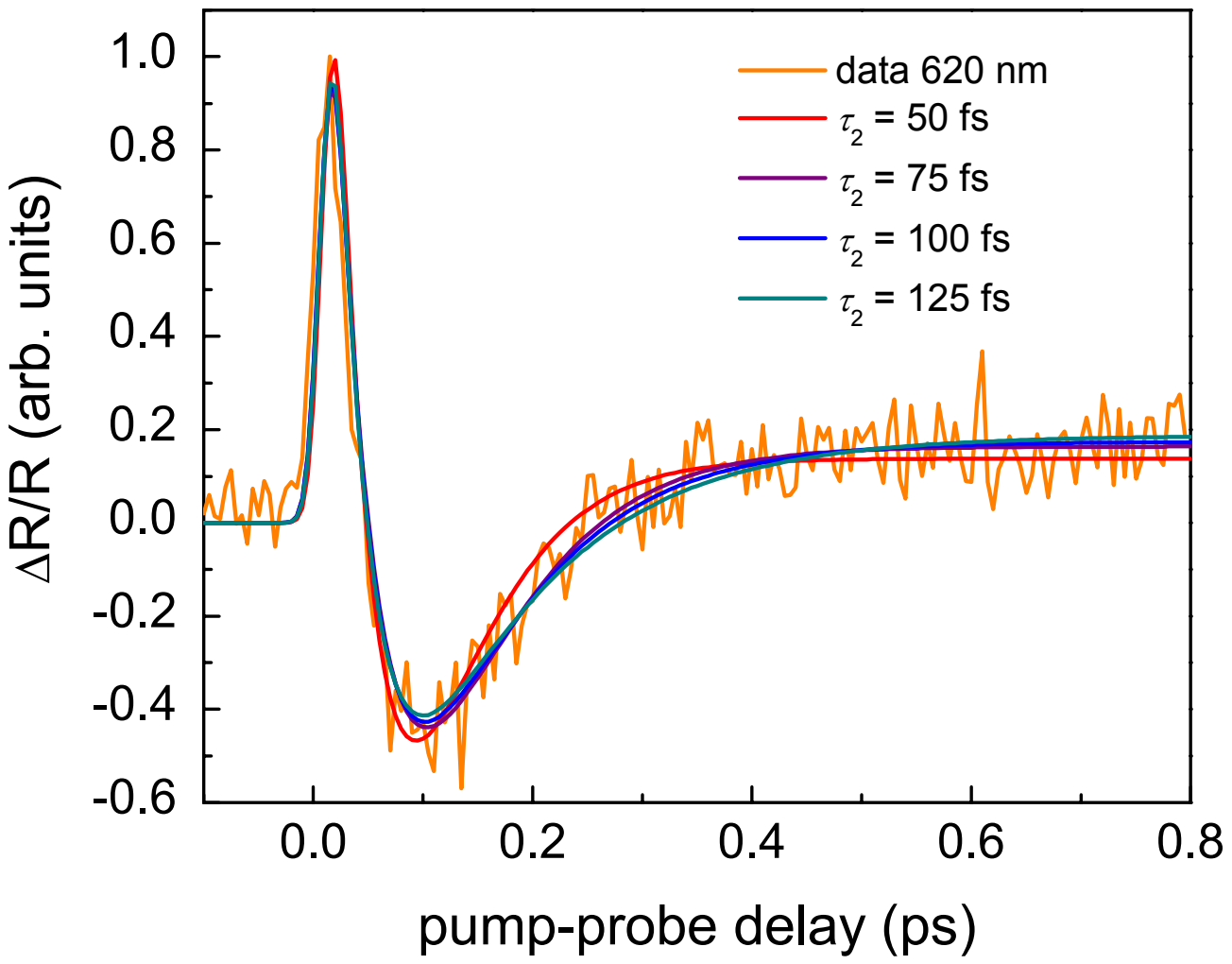

Figure S4. Normalized transient differential reflection of $Bi_2Sr_2CaCu_2O_{8.14}$ at 620 nm and best fits for different fixed values of $\tau_2$.

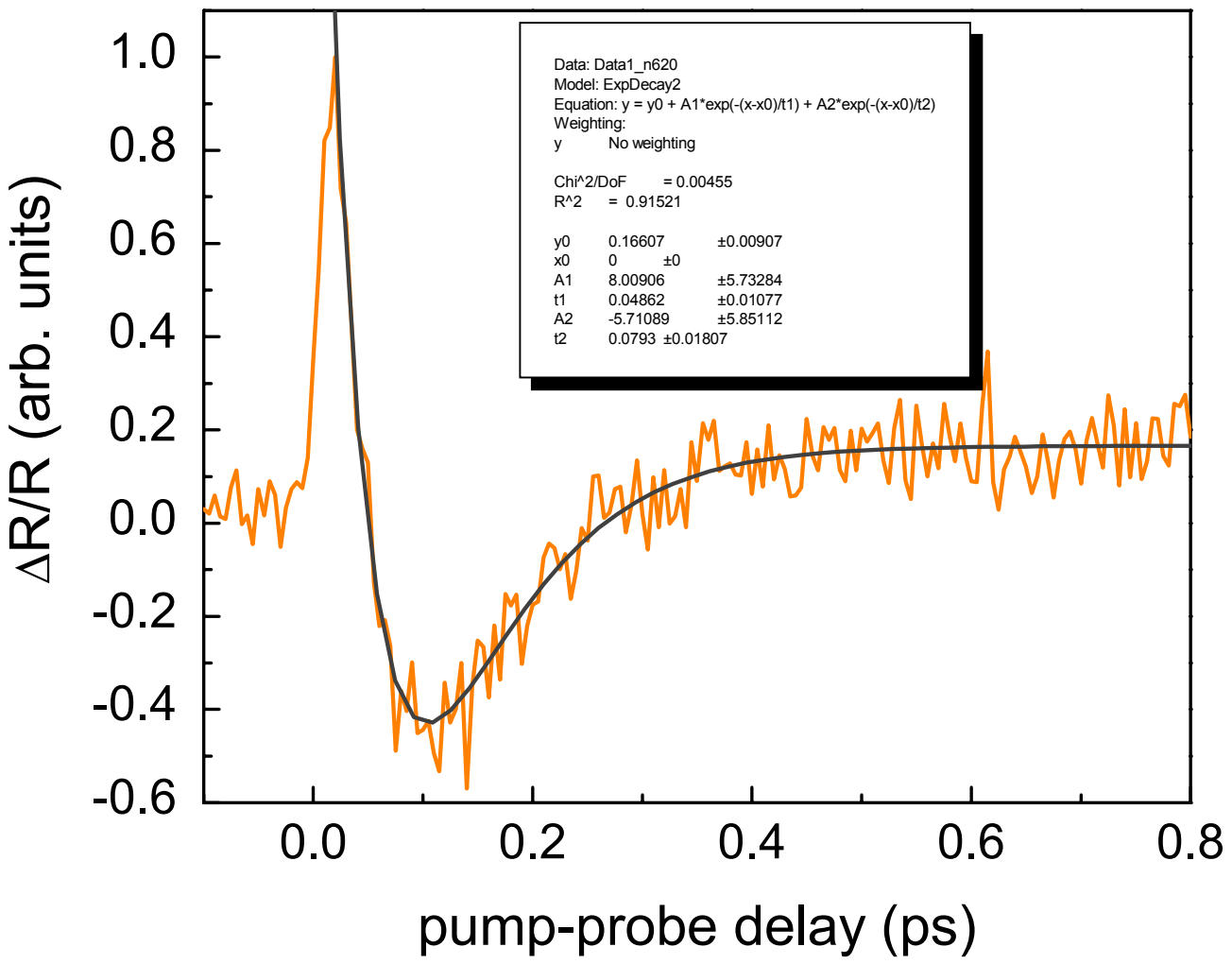

Figure S5. Normalized transient differential reflection of $Bi_2Sr_2CaCu_2O_{8.14}$ at 620 nm and fit to a double-exponential decay.

Finally, we assess the influence of the data's noise on our fits. Since, as mentioned above, our procedure does not yield error margins, we fitted the parts of our time traces (with the oscillating part subtracted where necessary) where the pump-probe overlap becomes negligible (i.e. for a pump-probe delay > 30 fs) as a simple sum of two exponential decays and a plateau. Any data analysis software gives the error margins of these fit parameters via a $\chi^2$-square analysis (we used OriginLab Origin 7.0). Figure S5 shows such a fit for $Bi_2Sr_2CaCu_2O_{8.14}$ at 620 nm, the same time trace that has been used in Figure S4 to demonstrate the cross-talk between $\tau_1$ and $\tau_2$. The uncertainty on $\tau_2$ is comparable to that assumed above, and the error margin for $\tau_1$ is 20%. Note that this margin contains both the influence of the noise and of the mutual influence of $\tau_1$ and $\tau_2$ on each other. We used this procedure to calculate the relative uncertainty of $\tau_1$ for all our data and list them in Table S1. For a large dataset on the same sample, the error margins obtained this way should coincide with those obtained from the variance of the $\tau_1$ values obtained from different time traces (different probe wavelengths and sample orientations) unless the latter is increased by additional small signal components we neglected. In Table S2 we report all $\Gamma_1$ values obtained from our fits using coupled differential equations, the relative uncertainty from their variance, and the relative uncertainty from quadratically summing the error margins from Table S1. For each sample the bigger of the two is printed in bold. In the last column, we use the bigger of the two and add an additional 4 fs error due to the uncertainty in the instrument resolution. This is the final error bar which we use in Figures 2 and 3 of the main paper. By far the highest uncertainties are found for $Bi_2Sr_2CaCu_2O_{8.14}$ and $La_{2-x}Sr_xCuO_4$, both because of the faster $\Gamma_1$ and the higher noise compared to the other data. In fact, their uncertainties are bigger than the difference between the values for $Bi_2Sr_2CaCu_2O_{8.14}$ and $La_{2-x}Sr_xCuO_4$, hence one could argue that they are undistinguishable. However, all other error margins are smaller and the sequence $YBa_2Cu_3O_{6+x}$ - $HgBa_2Ca_2Cu_3O_{8.2}$ - $Bi_2Sr_2CaCu_2O_{8.14}$, in order of ascending $\Gamma_1$, is without any doubt, which confirms that our data actually show a correlation between $a$ and $\Gamma_1$. As additional evidence we calculate Pearson's correlation coefficient, which is $r_c$ = -0.983 for the cuprates and $r_p$ = -0.987 for the pnictides, which corroborates or assumption of a linear relationship with a negative slope.

Table S1. Relative error in % for each individual $\Gamma_1$ value, final error as described in the text.

| Sample | 580 nm | | 620 nm | | 660 nm | 700 nm | total | final |
|---|---|---|---|---|---|---|---|---|
| $BaFe_2As_2$ | 3.1 | 5.3 | 5.2 | | | | **2.7** | **3.0** |
| $Ba(Fe_{0.975}Co_{0.025})_2As_2$ | 3.6 | 8.3 | 2.8 | 10 | 5.5 | 10.7 | 3.0 | **6.0** |
| $Ba(Fe_{0.949}Co_{0.051})_2As_2$ | 4.4 | 4.6 | 6.4 | | 7 | | 2.9 | **6.6** |
| $Ba(Fe_{0.93}Co_{0.07})_2As_2$ | 3.5 | | 6.1 | | 6.6 | | 3.2 | **11** |
| $Ba(Fe_{0.89}Co_{0.11})_2As_2$ | 3.7 | 7.8 | 7.2 | | | | 3.7 | **9.4** |
| $SmFeAsO_{0.8}F_{0.2}$ | | | 8.2 | | 6.6 | 7.7 | 4.3 | **18** |
| $YBa_2Cu_3O_{6.5}$ | 37 | | 43 | | 49 | | 25 | **28** |
| $YBa_2Cu_3O_{6.9}$ | 2.3 | 4.3 | 2.4 | | 2.9 | | 1.5 | **7.7** |
| $HgBa_2CuO_{4.1}$ | 11 | | 13 | | | | 8.5 | **13** |
| $Bi_2Sr_2CaCu_2O_{8.14}$ | 19 | | 22 | | | | **15** | **17** |
| $La_{1.9}Sr_{0.1}CuO_4$ | 93 | | 86 | 22 | | | **43** | **44** |
| $La_{1.85}Sr_{0.15}CuO_4$ | 38 | | 11 | | 18 | | 11 | **15** |

Table S2. $\Gamma_1$ value, their average and relative variance.

| Sample | 580 nm | | 620 nm | | 660 nm | 700 nm | average | Variance (%) |
|---|---|---|---|---|---|---|---|---|
| $BaFe_2As_2$ | 3.32 | 3.33 | 3.31 | | | | 3.2 | 0.3 |
| $Ba(Fe_{0.975}Co_{0.025})_2As_2$ | 3.2 | 3.29 | 3.5 | 3.2 | 3.1 | 10.7 | 3.2 | **5.9** |
| $Ba(Fe_{0.949}Co_{0.051})_2As_2$ | 2.9 | 3.18 | 3.1 | | 2.8 | | 3.0 | **6.5** |
| $Ba(Fe_{0.93}Co_{0.07})_2As_2$ | 3.1 | | 3.6 | | 2.9 | | 3.2 | **11** |
| $Ba(Fe_{0.89}Co_{0.11})_2As_2$ | 3.2 | 3.6 | 3 | | | | 3.3 | **9.4** |
| $SmFeAsO_{0.8}F_{0.2}$ | | | 4.3 | | 5.7 | 7.7 | 5.4 | **18** |
| $YBa_2Cu_3O_{6.5}$ | 11 | | 18 | | 12 | | 14 | **28** |
| $YBa_2Cu_3O_{6.9}$ | 12.0 | 12.2 | 13.2 | | 13.5 | | 12.7 | **5.8** |
| $HgBa_2CuO_{4.1}$ | 15.0 | | 17.5 | | | | 16.3 | **11** |
| $Bi_2Sr_2CaCu_2O_{8.14}$ | 19.4 | | 21.1 | | | | 20.3 | 5.9 |
| $La_{1.9}Sr_{0.1}CuO_4$ | 22.5 | | 26.6 | 23.8 | | | 24.3 | 8.6 |
| $La_{1.85}Sr_{0.15}CuO_4$ | 20.9 | | 22.4 | | 24.3 | 18.3 | 21.5 | **12** |

B. Interpretation of the observed relaxation behavior and implications for the high- *Tc* pairing mechanism

B1. Determining the EPI strength from the relaxation time

Our primary observation relates the dominant electron energy relaxation (EER) process to the structural parameter $a$. The EER can be related to the EPI using Allen's reformulation[48] of the two-temperature model (TTM)[48]:

$$\tau_E = \frac{\pi k_B}{3\hbar} \frac{T_e}{\lambda\langle\omega^2\rangle} \tag{S1}$$

where $\lambda<\omega^2>$ is the second moment of the Eliashberg spectral function, $T_e$ is the initial electronic temperature and $\tau_E$ is the energy relaxation time. The factor $\lambda<\omega^2>$ is of fundamental interest, as it quantitatively describes the weighted coupling of electrons to the lattice[48,48]. (In traditional electron-phonon models of superconductivity, it is directly related to $T_c$). The model was used in numerous studies on conventional and unconventional superconductors to determine $\lambda<\omega^2>$, with reasonable agreement, particularly in conventional metallic superconductors[Error! Bookmark not defined.] giving values between $\lambda = 0.08$ (for Cu) and $\lambda = 1.45$ (for Pb).

The main assumption of Allen's model was that the photoexcited electron (and hole) distributions thermalize on a timescale $\tau_{e-e}$ which is much faster than the electron-phonon relaxation rate $\tau_{e-ph}$, which was justified with the statement that "experimental monitors are so far consistent with the notion that even if the electron and phonon distribution functions are athermal, they are still adequately characterized by a thermal distribution at equivalent energy". However, recent time-resolved ARPES experiments[48] showed that the hot electron distribution departs from the thermal one, particularly at high energies.

Allen's TTM made one additional prediction: $\tau_{e-ph}$ should be strongly dependent on $T_e$, and thus also on the pump laser intensity, but this important dependence was never confirmed by experiments, suggesting that the model's applicability to the high-temperature superconductors may need to be reexamined.

Recently Kabanov and Alexandrov (2008) suggested that the original assumption that $\tau_{e\text{-}e} << \tau_{e\text{-}ph}$ may not be valid, and obtained an exact analytical solution to the Boltzmann equations without making this a-priori assumption[48]. They calculated the e-e scattering rate time, which for a typical high- $T_c$ material gives $\tau_{e\text{-}e} \sim 1$ ps (here we have used $\mu_c \sim 1$ and $E_F = 0.5$ eV), which is much longer than the measured relaxation times. Their expression for the electron phonon energy relaxation time is very similar to Allen's formula, with the exception of a numerical factor of 2, and the important replacement of $T_e$ by the lattice temperature $T_L$:

$$\tau_E = \frac{2\pi k_B}{3\hbar} \frac{T_L}{\lambda\langle\omega^2\rangle} \qquad \text{(S2)}$$

The theory eliminates the problem of determining the initial electron temperature, opening the way to systematic direct and accurate measurements of $\lambda<\omega^2>$, since $T_L$ can be accurately determined. Importantly, the model also predicts that $\tau_{e\text{-}ph}$ should be independent of pump laser intensity, which is eminently verifiable experimentally. Making a rough estimate of the expected relaxation time, using Eq. S2, the predicted $\tau_{e\text{-}ph}$ = 40-400 fs, depending on $\lambda<\omega^2>$ indicates that experimental resolution beyond the 70 fs used so far is required to systematically investigate the EPI. Importantly, the model applies to the high-temperature state, which is assumed to be described by Boltzmann's equation, with electron correlations taken into account with a pseudopotential. Thus, experiments measuring $\lambda<\omega^2>$ need to be performed at high temperatures in order to avoid correlation effects. Note that the Eqs S1 and S2 differ only by a factor of 2 and, more importantly, the fact that the EER is given in terms of the accurately measurable $T_L$ rather than $T_e$.

To decide whether the TTM is an acceptable approximation, we recently determined $\tau_{e\text{-}ph}$ for YBCO and LaSCO with pump intensities in the range 50 – 500 μJcm$^{-2}$, which significantly heat the electron system but not the lattice[48]. We estimated effective electron temperatures after excitation between 400 and 800 K, depending on the excitation density. However, the measured $\tau_{e\text{-}ph}$ showed no dependence on the pump intensity. Hence while Equation S2 reliably yields the same $\lambda<\omega^2>$ for data taken at different excitation densities, the TTM estimates depend on the inferred initial electron temperature after excitation. This shows that the consequences of the inappropriate TTM assumption in Allen's model $\tau_{e\text{-}e} << \tau_{e\text{-}ph}$ are not limited to a simple numerical correction factor, but lead to qualitative disagreement with experiment.

To assess the validity of Equation 2 we experimentally verified the prediction $\tau_{e\text{-}ph} \sim T_L$, as stated in Equation 2. We observed the relaxation behavior of $La_{1.85}Sr_{0.15}CuO_4$ – representing the samples with the lowest $\tau_{e\text{-}ph}$ within our study – in the temperature range 140 – 310 K and found a linear increase of $\tau_{e\text{-}ph}$ with $T_L$ (see Fig S5). For lower temperatures the spectral weight of the fast component becomes too low to be fitted reliably (7% at 140 K and indiscernible for lower temperatures). A linear increase of $\tau_{e\text{-}ph}$ with $T_L$ has been observed also for $SmFeAsO_{0.8}F_{0.2}$[48]. Note, however, that in that case the linear $T_L$ dependence shows only above 230 K. Below this temperature the signal is dominated by the pseudogap component, which has a maximum $\tau$ at the pseudogap temperature and falls off on both sides. A single-exponential fit can even lead to an artificial decrease of $\tau$ with $T_L$, when the spectral weight of the faster $\tau_1$ increases relative to $\tau_2$ with increasing $T_L$. Also in Figure S5 this increasing spectral weight of $\tau_1$ gives the illusion of a decay that becomes faster with increasing temperature, when in fact both components become slower.

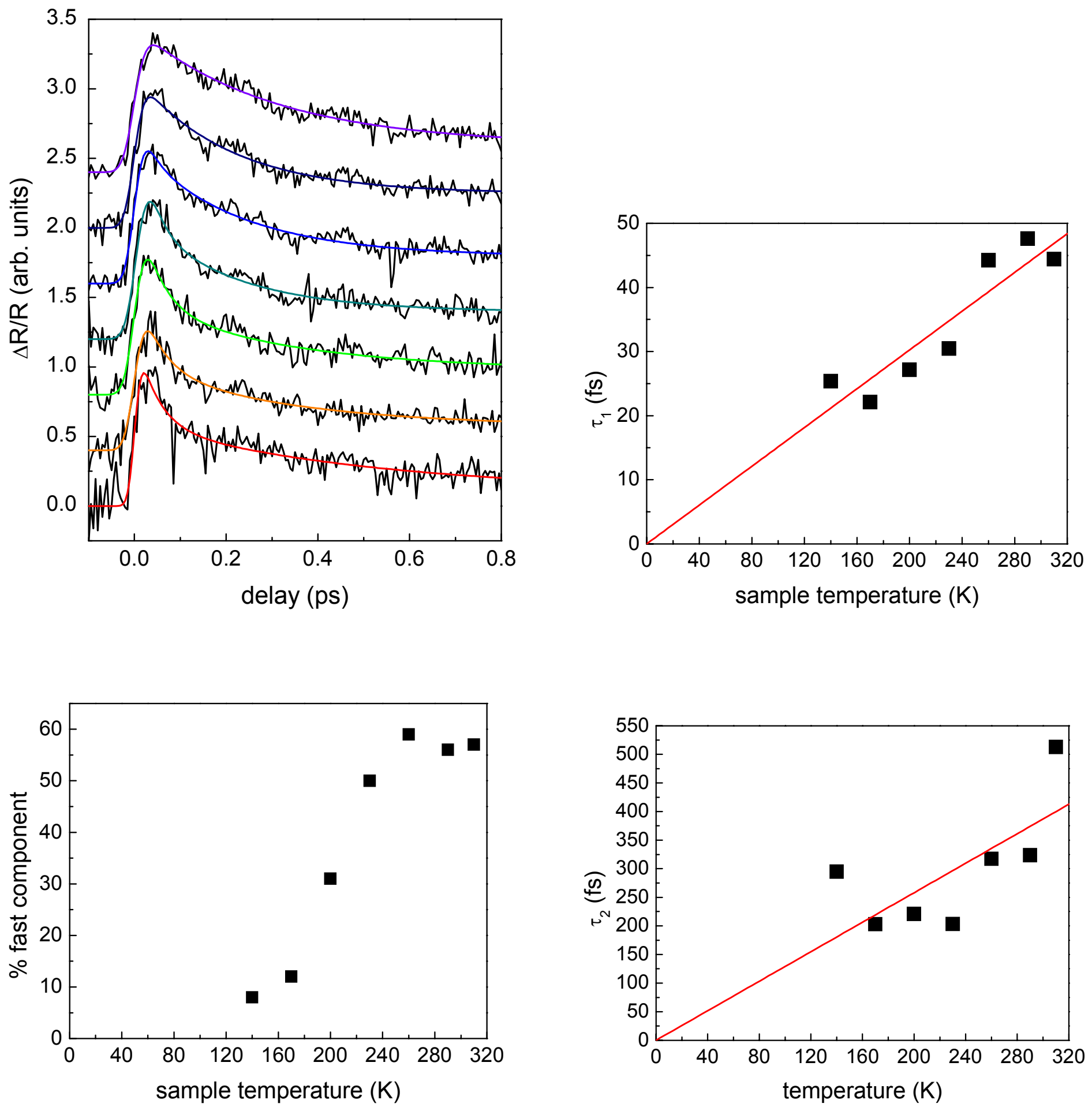

Figure S6. Top left: Normalized transient differential reflection at 580 nm of La$_{1.85}$Sr$_{0.15}$CuO$_4$ for temperatures from 140 to 310 K (top to bottom) fitted with two exponential decays. Bottom left: Spectral weight of the fast component as a percentage of the signal. Right: The energy relaxation times of La$_{1.85}$Sr$_{0.15}$CuO as a function of the sample temperature.

All theoretical descriptions of the EER that yield a directly applicable relationship between $\tau_{e\text{-}ph}$ and the EPI strength (such as Eqs S1 and S2), to date, have used two concepts from the Fermi liquid (FL) theory: the notion of quasiparticles and the description of their relaxation via collision integral. Quasiparticles are well defined if their energy uncertainty due to lifetime broadening is small compared to their excitation energy $E_e$. The quasiparticle lifetime is given by

$$\tau_e(E_e) = \frac{2}{K\left(\pi^2 k_B^2 T^2 + E_e^2\right)} \tag{S3}$$

With the estimate for $K$ given by:

$$K \approx \frac{a^2}{8\pi\hbar E_F a_B^2} \tag{S4}$$

With $E_F$ being the Fermi energy, $a$ the mean electron distance and $a_B$ the Bohr radius.

Hence the relative lifetime broadening of the quasiparticle energy is:

$$\frac{\Delta E_e}{E_e} = \frac{\hbar}{\tau_e E_e} \approx \frac{\hbar K E_e}{2} \approx \frac{E_e a^2}{16 E_F a_B^2} \tag{S5}$$

Where we assumed $E_e >> \pi k_B T$ in the second step. Therefore $\Delta E_e/E_e$ decreases during the EER. The upper bound for $E_e$ is the exciting photon energy of 2.3 eV at $t = 0$. For an exciting photon energy of 2.3 eV and a Fermi energy of $E_F = 0.5$ eV, $\Delta E_e/E_e \sim 0.3$ at $t = 0$ and quasiparticles are well defined during the whole relaxation process. For underdoped samples with smaller Fermi energies the quasiparticle concept may be at the limit for very short delays $t < \tau_{e\text{-}ph}$, but is still reliable for all longer delays.

For the validity of the collision integral treatment there is a clear cut condition[15]: $\hbar\omega \lesssim E_\mathrm{F}$. In our case $E_\mathrm{F}$ is about one order of magnitude higher than the highest relevant phonon energy. Hence, the deviations from FL behavior observed in some experiments do not invalidate the quasiparticle quasiparticle concept and the collision integral formalism, they only show that for these cases additional corrections originating from electron correlation and or impurities may need to be considered. Since Eq 2 correctly predicts the experimentally observed (see next paragraph) linear dependence of $\tau_{e\text{-}ph}$ on the sample temperature and its independence of the pump intensity, such departure from pure FL behavior cannot yield a relation that is qualitatively different from Equation 2; it can only appear as a prefactor. Such a prefactor may depend on the strength of electronic correlations and hence on the doping level. However, within one compound family $1/\tau_{e\text{-}ph}$ varies only moderately with different doping levels, and this variation is fully explained by doping dependent strain and the concomitant changes in $\lambda\langle\omega^2\rangle$. Therefore, $1/\tau_{e\text{-}ph}$ is an appropriate measure for comparing the EPI strength of our compounds.

Finally, we emphasize that irrespective of which model is used to ultimately connect $\tau_E$ with the EPI, our presented systematics relate three experimentally measured quantities, $\tau_E$, $a$, and $T_c$ without prejudice as to the exact model relating $\tau_E$ to the microscopic EPI strength. The identification of the EPI as the dominant EER mechanism relies on the following arguments: (i) the unanimous consensus in the literature (See Refs 8-14, 26-28 in the main text. Note that Ref 11 invokes a highly complex relaxation scheme with four relaxation processes including a spin fluctuation channel. However, our data do not support a reliable extraction of so many fit parameters, therefore we limit our discussion to the dominant relaxation mechanism), (ii) only the phonon bath has sufficient heat capacity[14], (iii) the experimental confirmation of the fluence independence and linear temperature dependence of the primary relaxation time $\tau_1$ (Figure S5 and Ref 14 in the main text), (iv) the absence of a faster component, which would be visible as a resolution-limited initial spike if it gave a significant signal contribution, (v) the chain of reasoning in our paper which relates the relaxation time $\tau_1$ to $a$ and $a$ to the EPI, hence $\tau_1$ should be related to the EPI as well, and (vi) the perfect agreement of the measured $\tau_1$ with the expected relaxation times obtained using Equation S2 and estimates of $\lambda\langle\omega^2\rangle$ from other measurements (Refs 20, 23, 28, 33-38 in the main text).

B2 the second relaxation time $\tau_2$

The data for the pnictide samples can be fitted very well with a single-exponential decay and a plateau, plus one coherent oscillation. Hence, over the observed temporal window up to 800 fs after excitation, no appreciable second decay process could be identified. For the cuprates and bismuthates, a second decay component is found. Its dependence on the sample temperature, as measured for $La_{1.85}Sr_{0.15}CuO$ (see Fig S5, bottom right panel), suggests that it also obeys a relation analogous to Equation S2. However, as shown in Figure S6, there is no apparent correlation between the time constant of this relaxation and $T_c$. Hence this relaxation pathway, whose assignment is beyond the scope of this paper, does not seem to be important in determining $T_c$.

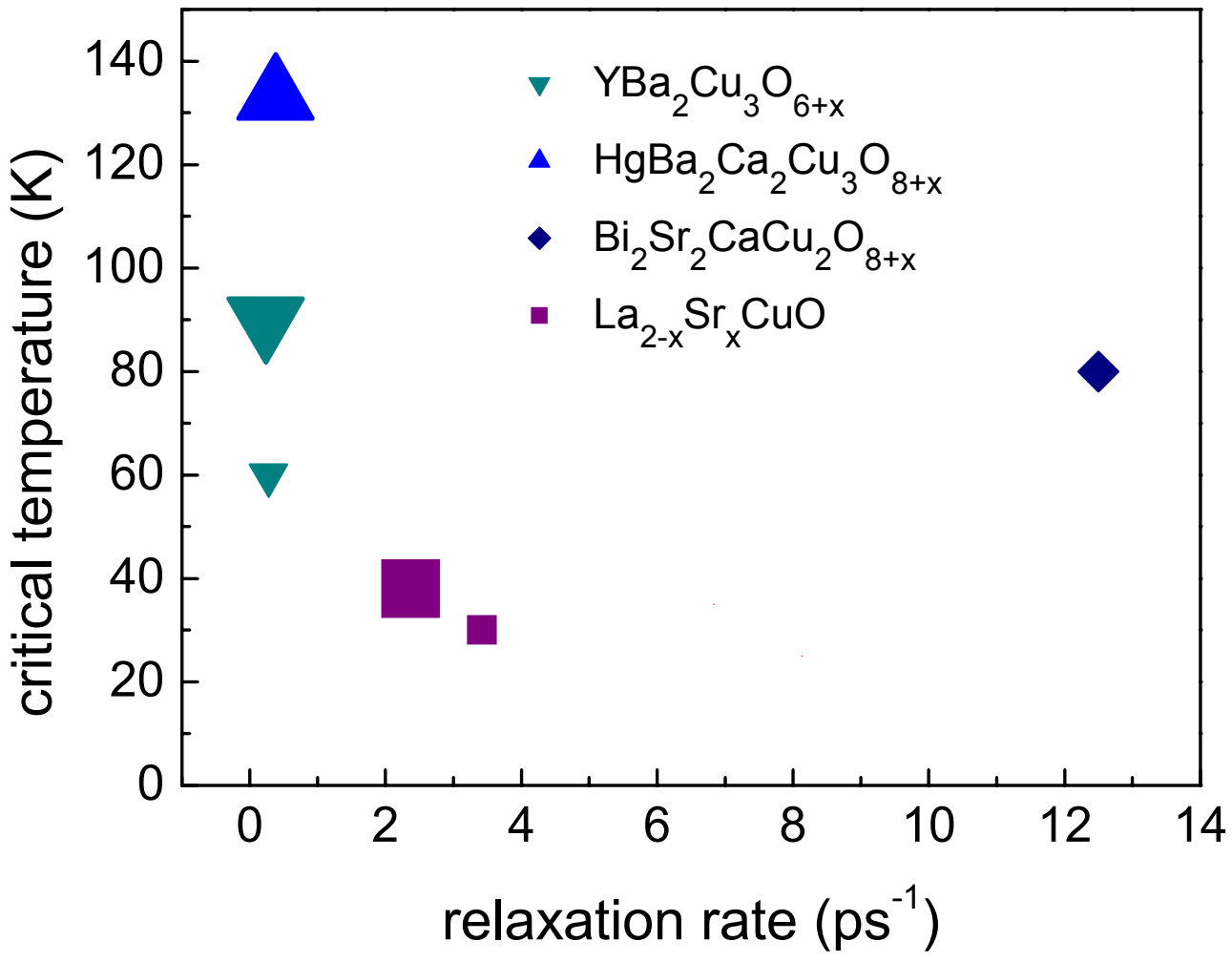

Figure S7 Superconducting critical temperatures as a function of the second fastest relaxation time.

C. Raw data and fits

In the following we display the data for all samples, normalized and fitted as described in section A4. Probe wavelengths are indicated in the figures; where the signal quality was sufficient, two orthogonal sample orientations were measured to average out possible anisotropies.

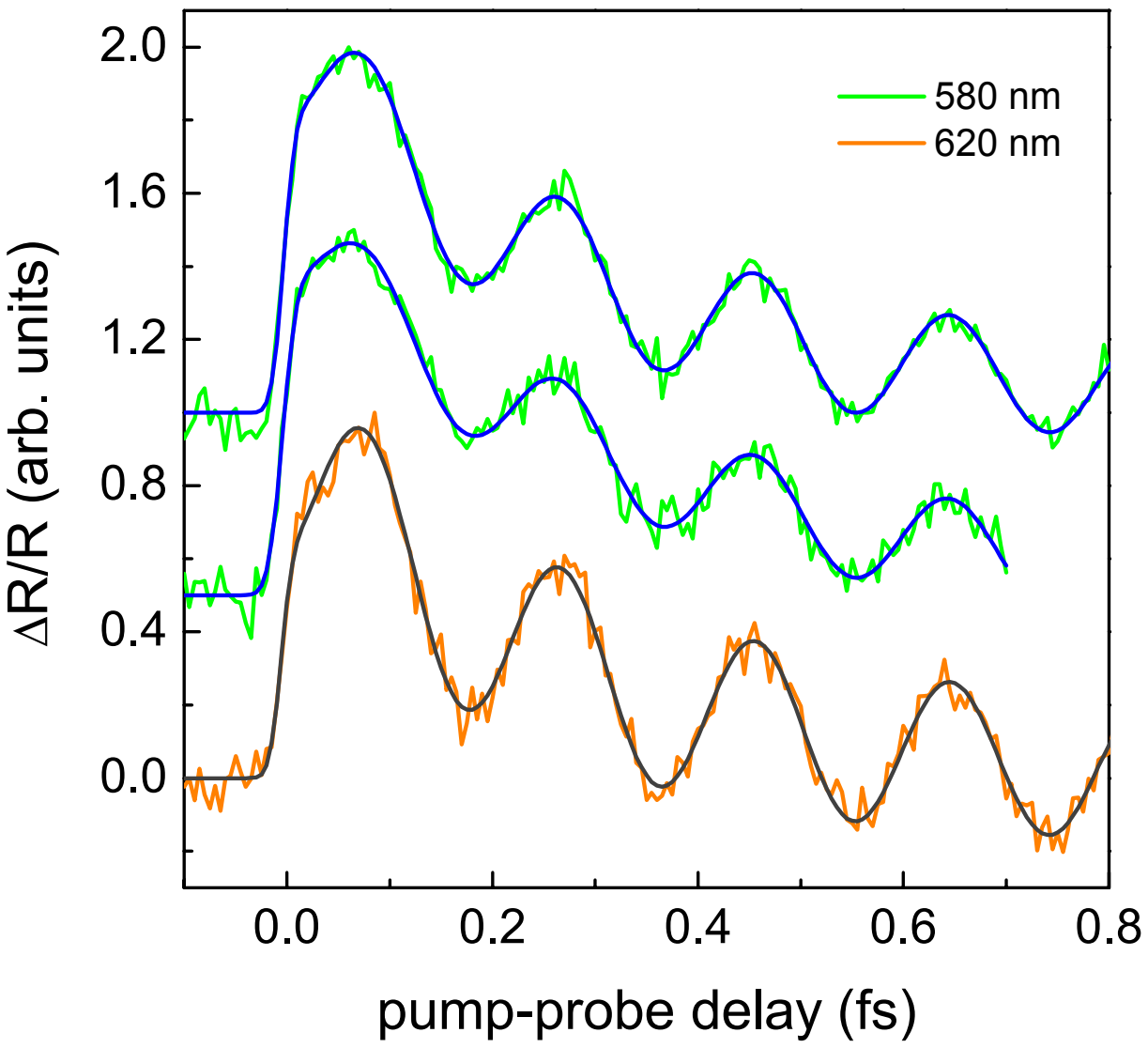

Figure S8. Normalized transient differential reflection of $BaFe_2As_2$.

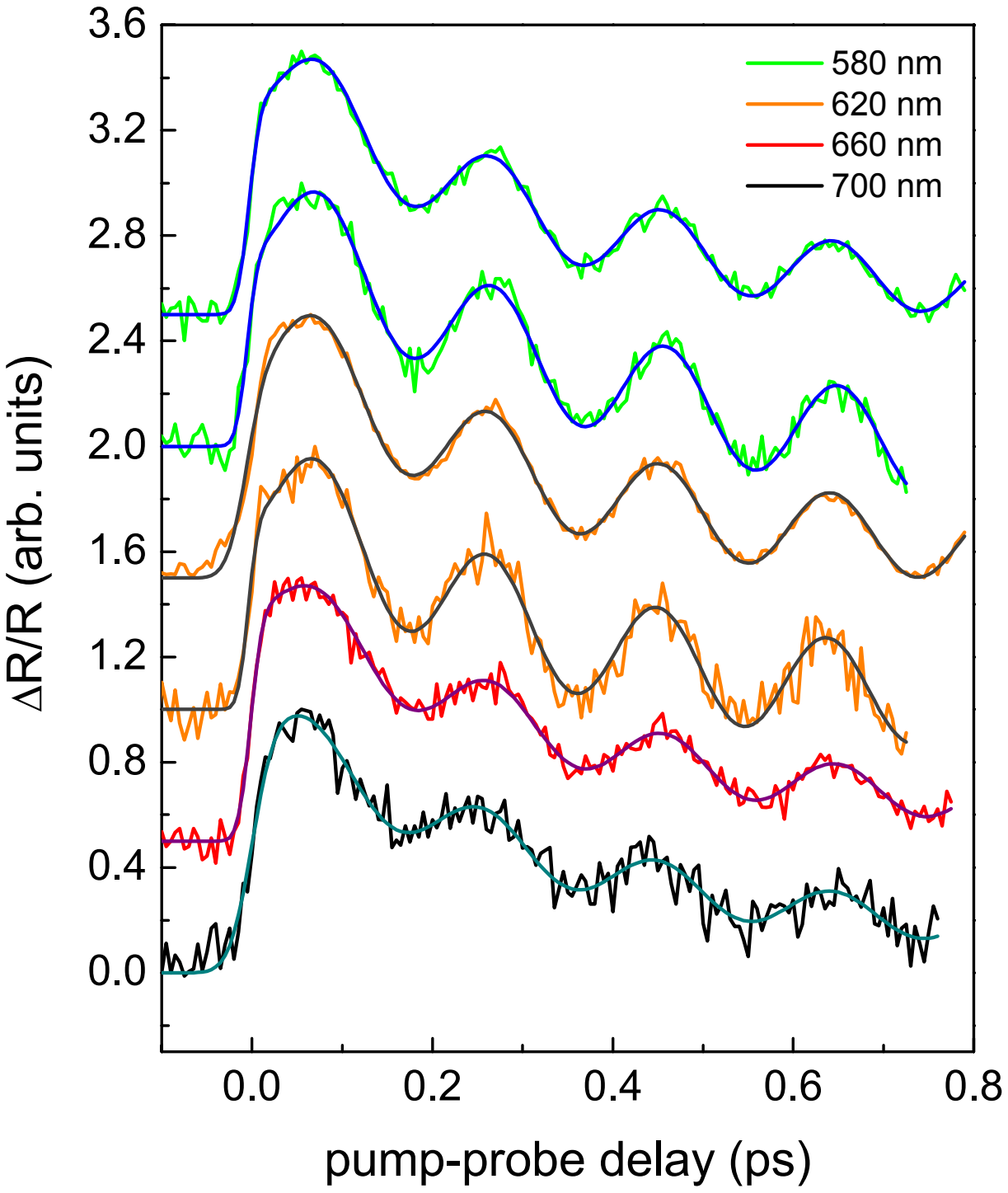

Figure S9. Normalized transient differential reflection of $Ba(Fe_{0.975}Co_{0.025})_2As_2$.

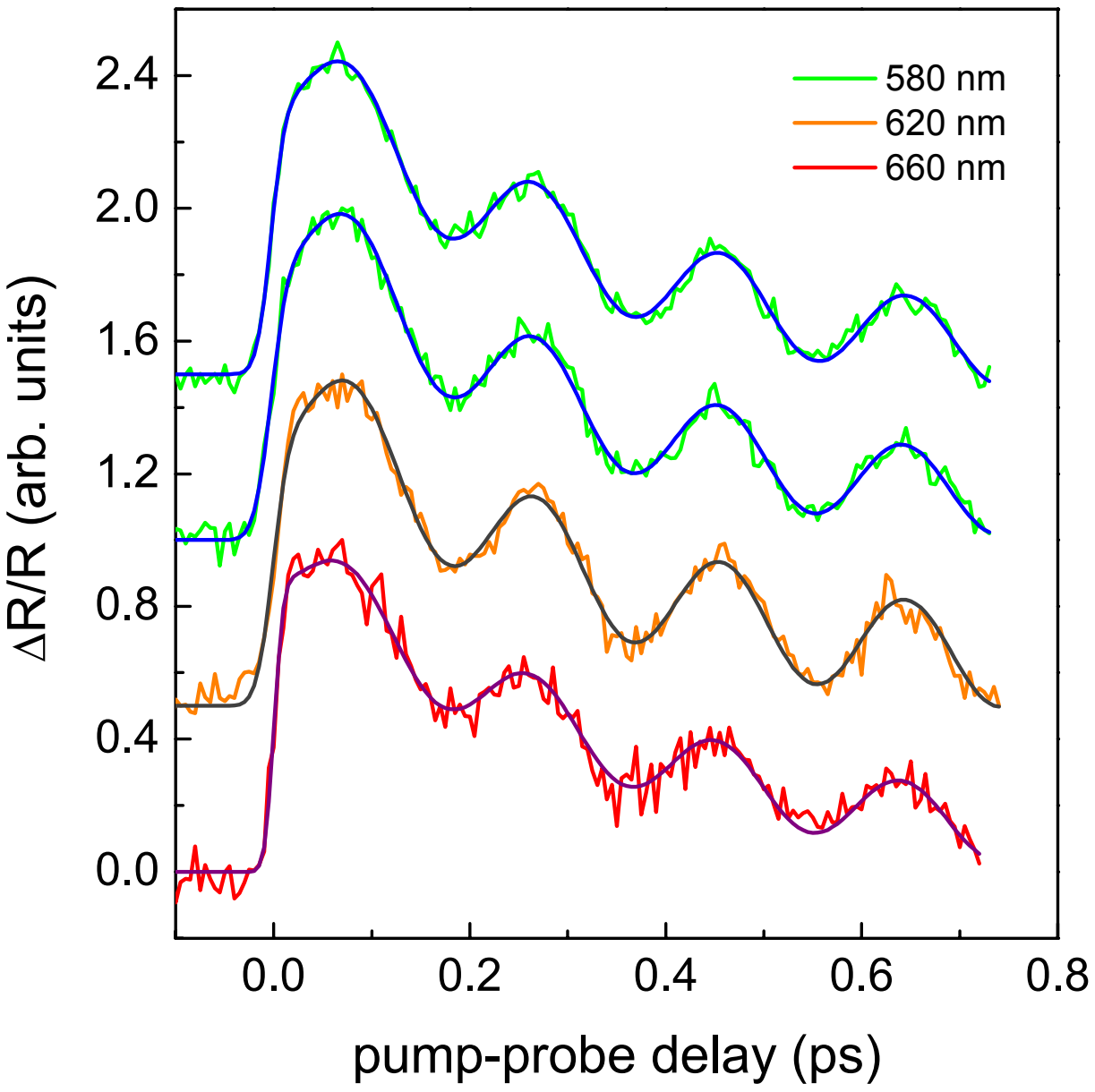

Figure S10. Normalized transient differential reflection of $Ba(Fe_{0.949}Co_{0.051})_2As_2$.

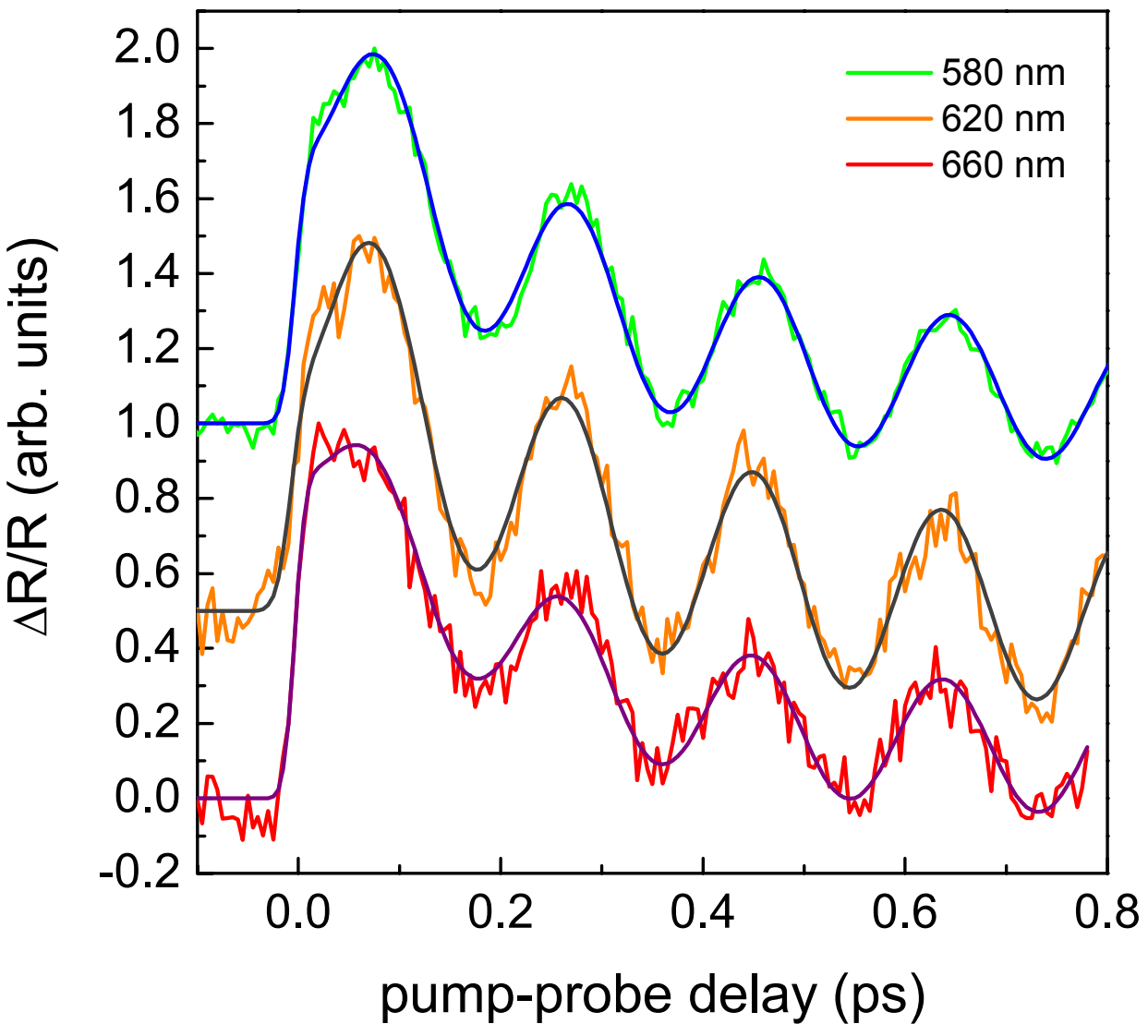

Figure S11. Normalized transient differential reflection of $Ba(Fe_{0.93}Co_{0.07})_2As_2$.

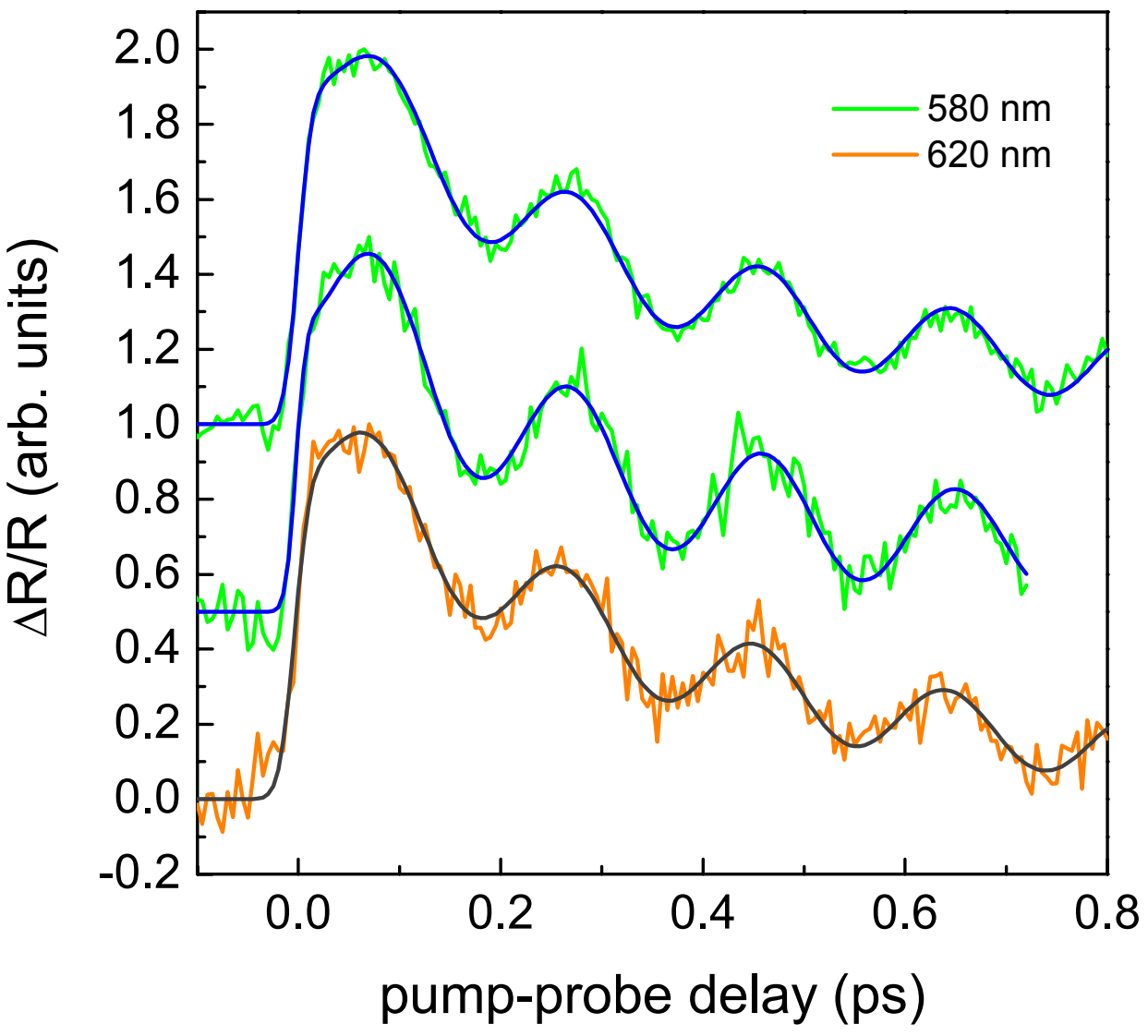

Figure S12. Normalized transient differential reflection of $Ba(Fe_{0.89}Co_{0.11})_2As_2$.

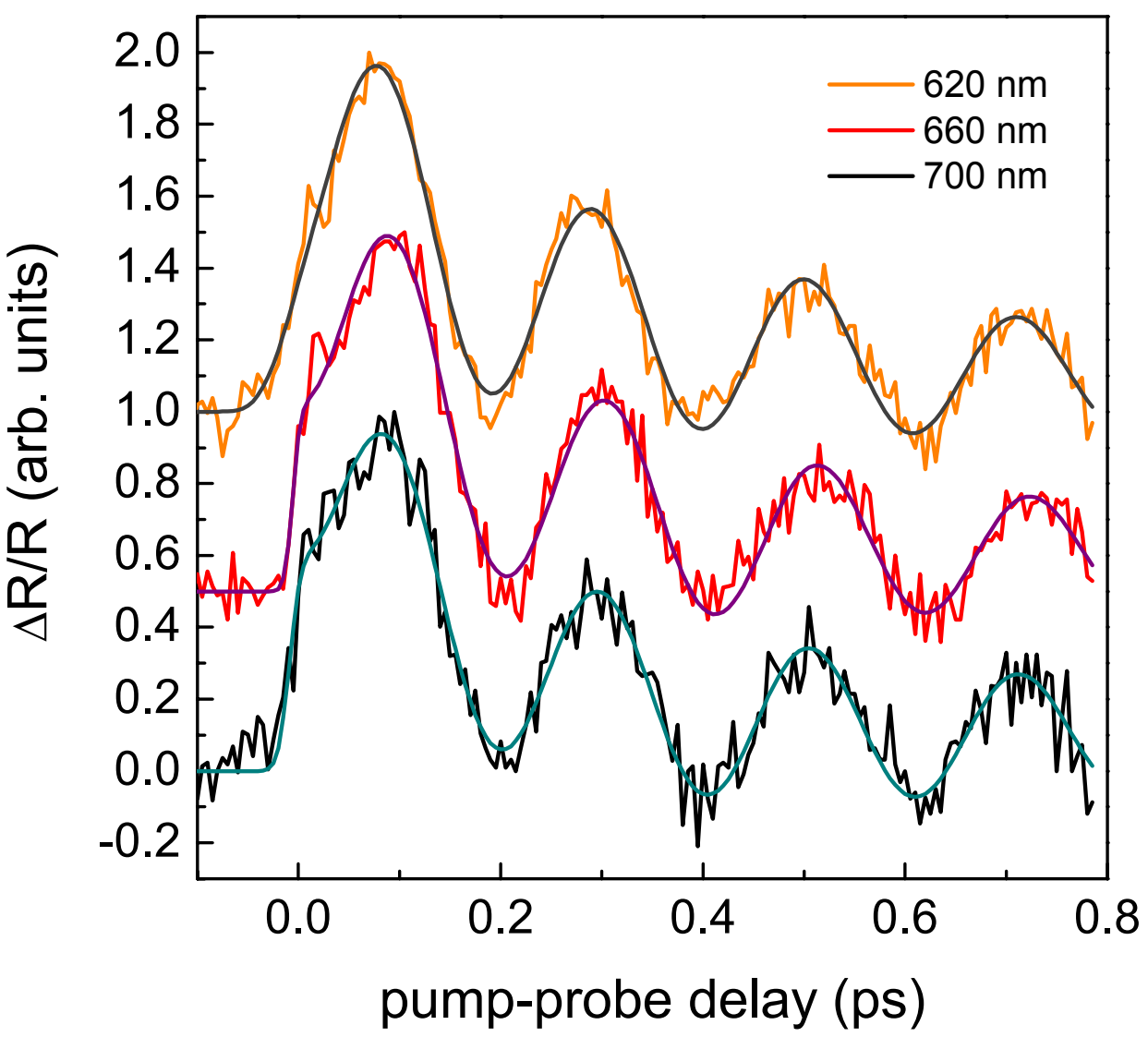

Figure S13. Normalized transient differential reflection of $SmFeAsO_{0.8}F_{0.2}$.

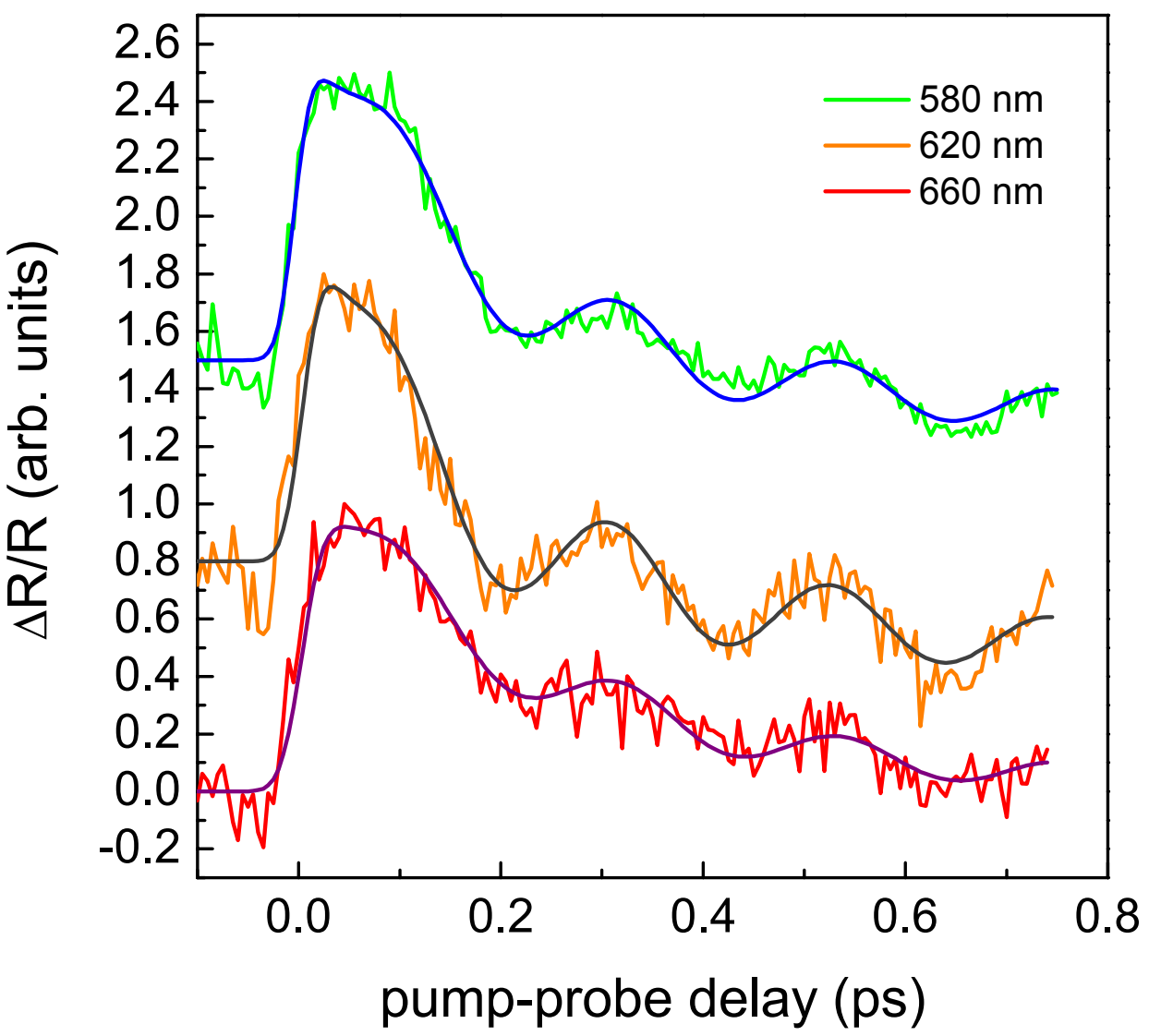

Figure S14. Normalized transient differential reflection of $YBa_2Cu_3O_{6.5}$.

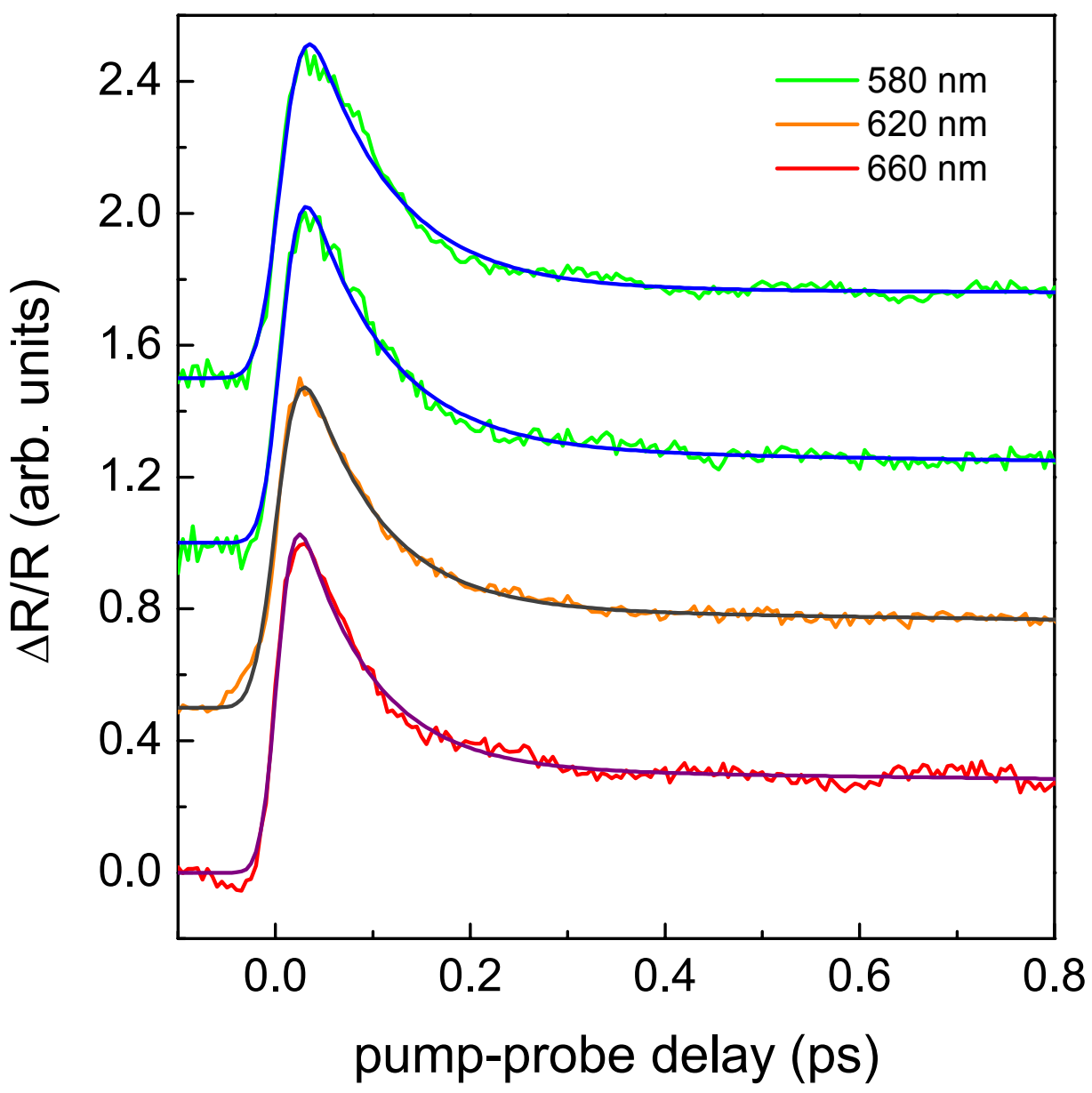

Figure S15. Normalized transient differential reflection of $YBa_2Cu_3O_{6.9}$.

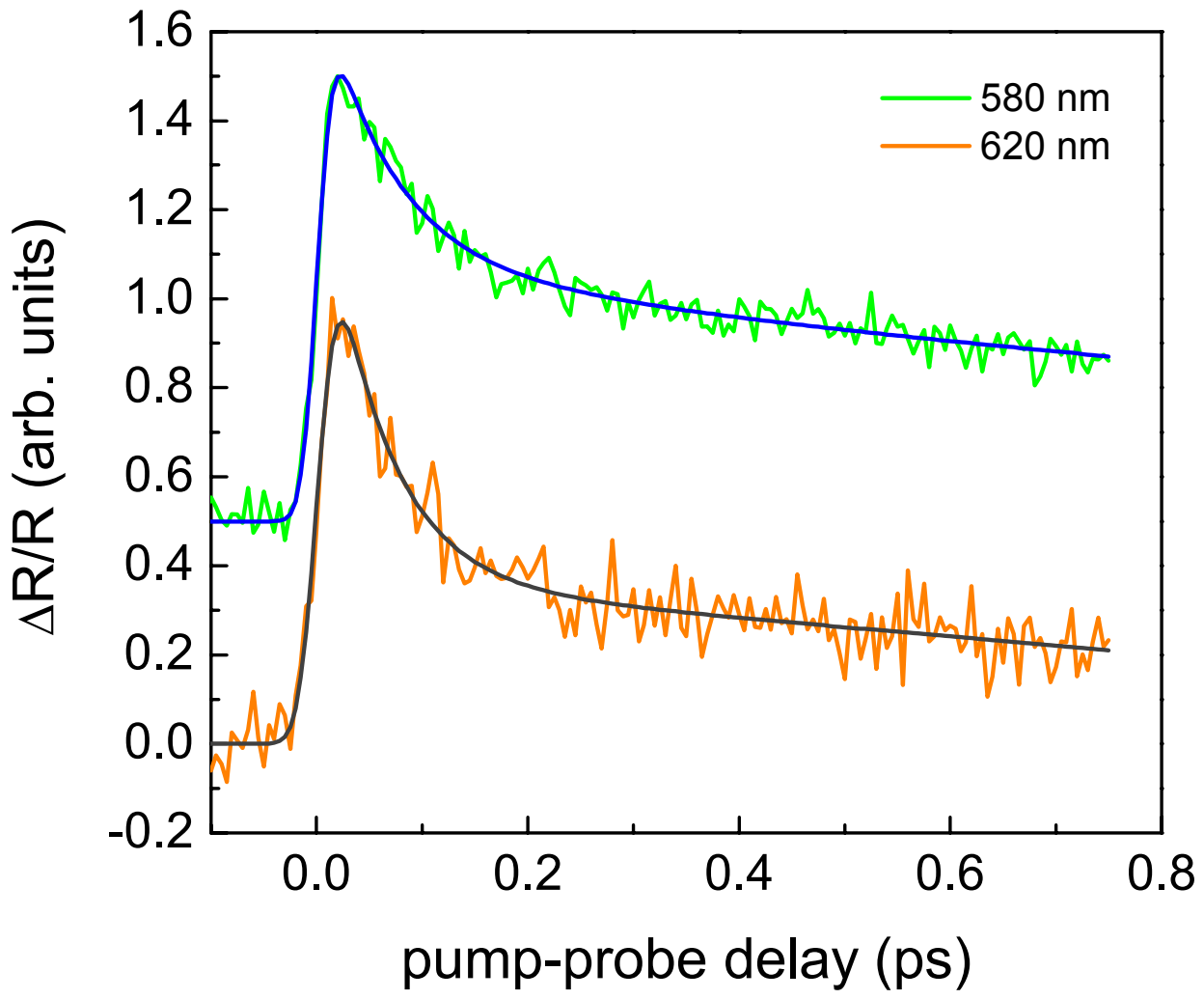

Figure S16. Normalized transient differential reflection of $HgBa_2Ca_2Cu_3O_{8.2}$.

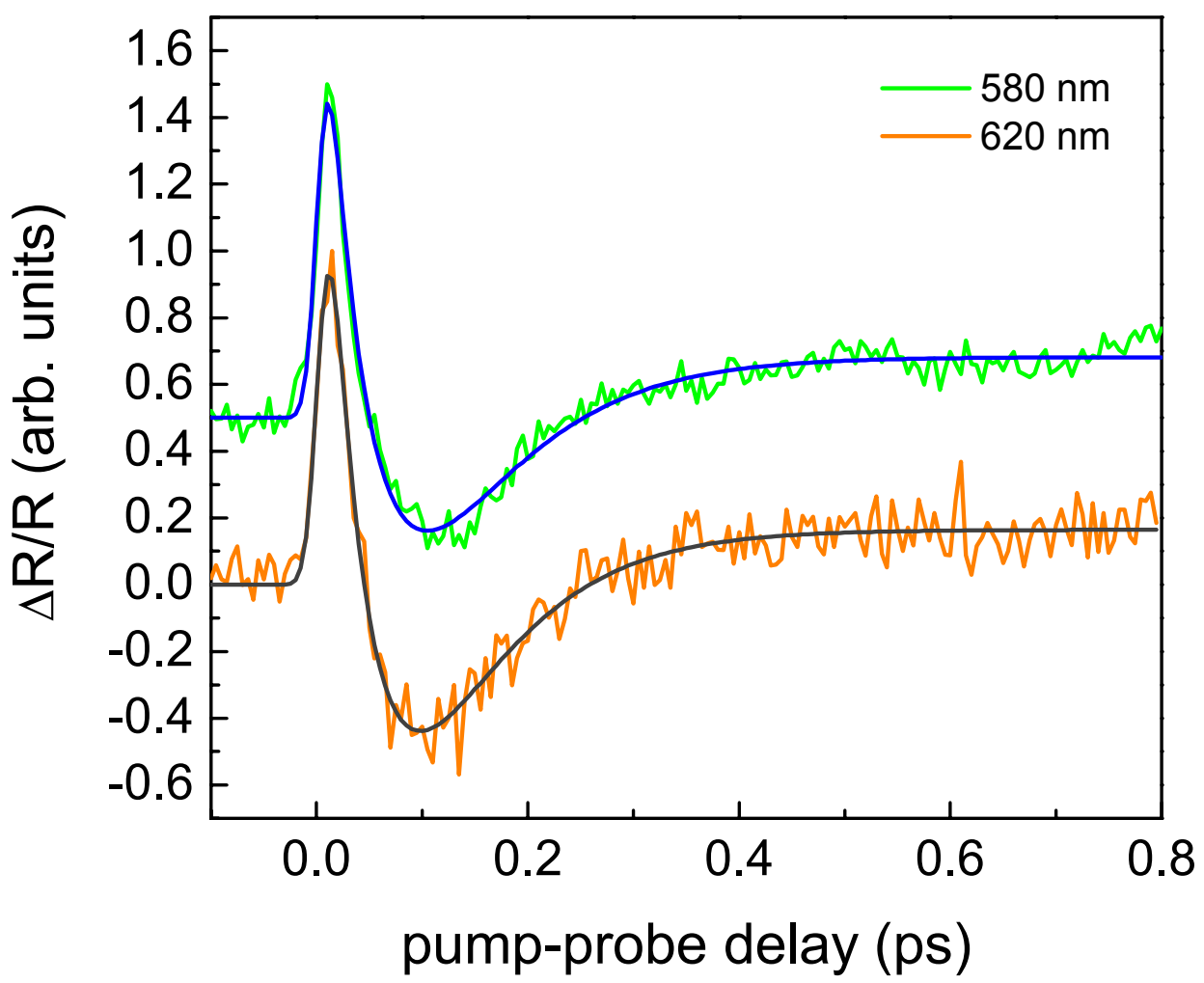

Figure S17. Normalized transient differential reflection of $Bi_2Sr_2CaCu_2O_{8.14}$.

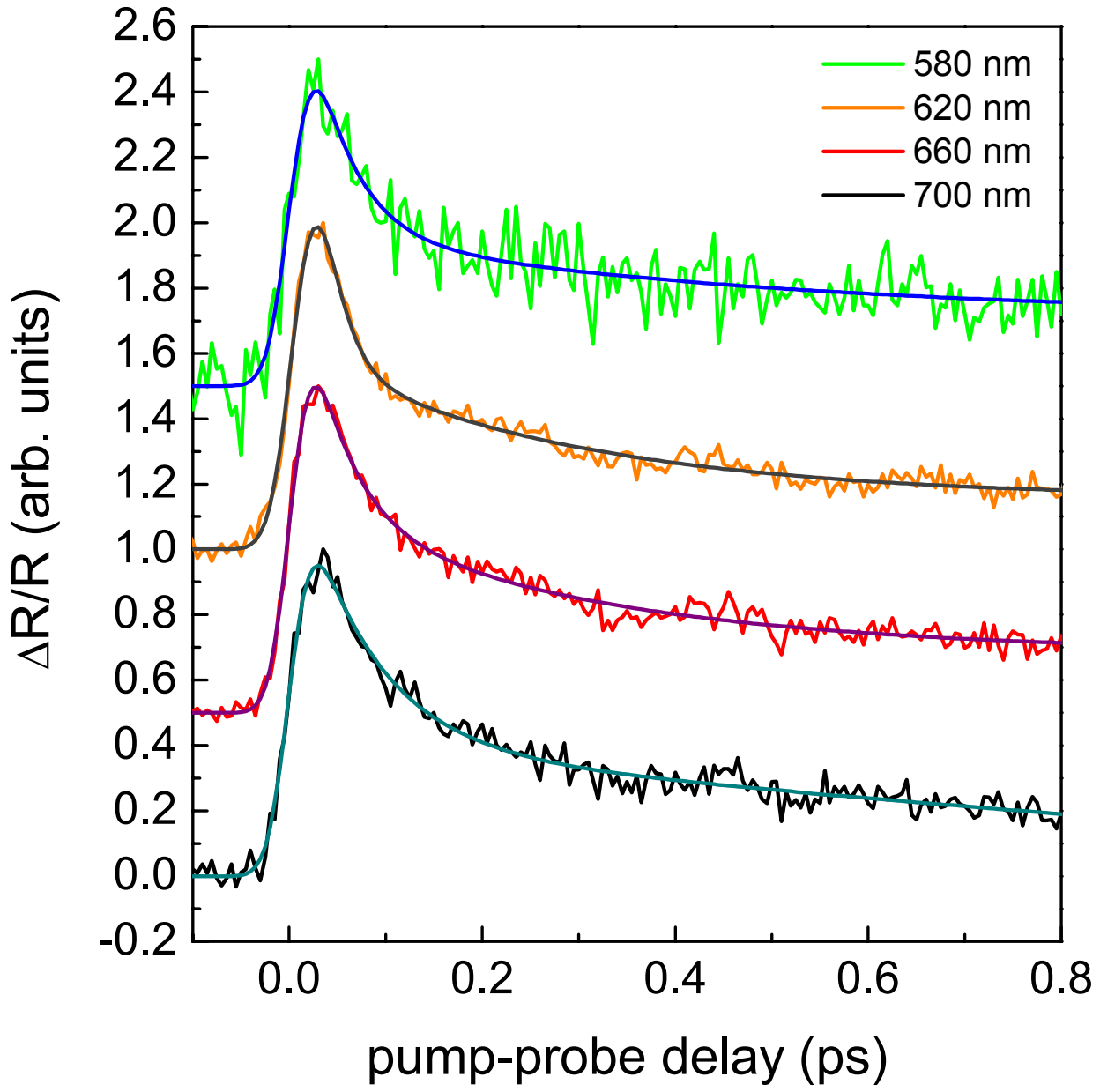

Figure S18. Normalized transient differential reflection of $La_{1.85}Sr_{0.15}CuO_4$.

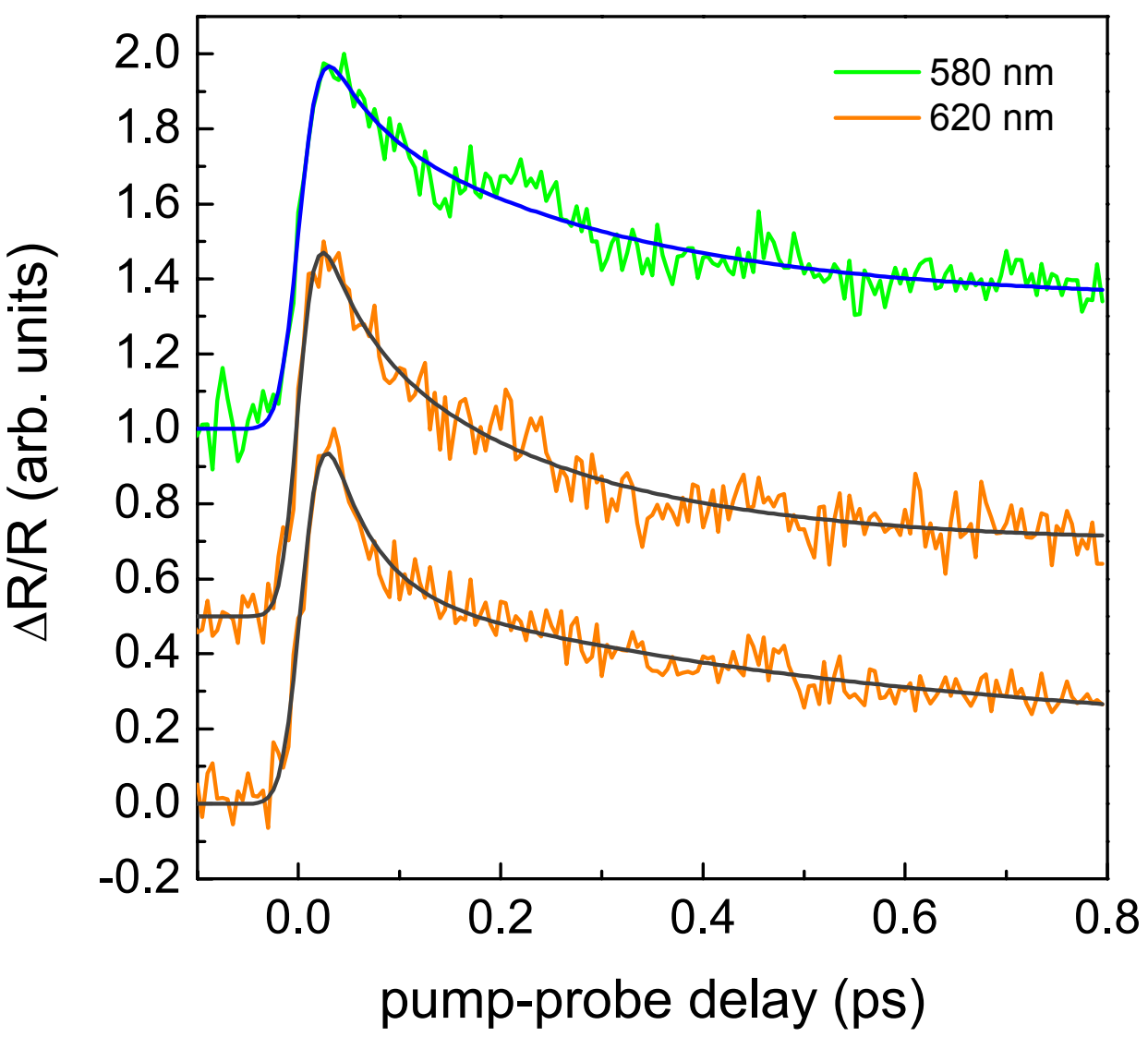

Figure S19. Normalized transient differential reflection of $La_{1.9}Sr_{0.1}CuO_4$.